\documentclass[11pt]{scrartcl} 
\usepackage{multicol}
\usepackage{wrapfig}

\usepackage{amsmath,amssymb,latexsym,amsopn,amsfonts}
\usepackage[boxed,ruled,vlined,linesnumbered,noresetcount]{algorithm2e}
\newcommand\bull{{\operatorname{-\xspace}}}
\usepackage{framed,microtype}
\usepackage[normalem]{ulem}
\usepackage{array,xspace}
\usepackage{graphicx}
\usepackage{color}
\newcommand{\ems}[1]{{#1}} 
\definecolor{oskar_green}{rgb}{0.0, 0.5, 0.0}

\newcommand{\algSize}{normalsize} 
\newcommand{\algSizeSmall}{scriptsize} 

\newcommand{\B}{}
\newcommand{\BB}{}
\newcommand{\BBB}{}

\newcommand{\bigO}{\mathcal{O}\xspace}
\newcommand{\remove}[1]{}
\newcommand{\reduce}[1]{#1}

\newcommand{\txD}{\mathit{txDes}\xspace} 
\newcommand{\exist}{\mathsf{hasTerminated}\xspace}

\newcommand{\done}{\mathsf{result}\xspace}
\newcommand{\deactivate}{\mathsf{deactivate}\xspace}
\newcommand{\test}{\mathsf{test}\xspace} 
\newcommand{\dCS}{\bot} 
\newcommand{\dBS}{\bot} 
\newcommand{\counts}{\mathsf{counts}\xspace}
\newcommand{\checkC}{\mathsf{check}\xspace}

\newcommand{\etal}{\emph{et al.}\xspace}
\newcommand{\eg}{\emph{e.g.,}\xspace}
\newcommand{\Eg}{\emph{E.g.,}\xspace}
\newcommand{\ie}{\emph{i.e.,}\xspace}
\newcommand{\Ie}{\emph{I.e.,}\xspace}

\newtheorem{remark}{Remark}[section]
\newtheorem{theorem}{Theorem}[section]
\newtheorem{lemma}[theorem]{Lemma}
\newtheorem{definition}{Definition}[section]

\newtheorem{assumption}[theorem]{Assumption}
\newtheorem{corollary}[theorem]{Corollary}
\newtheorem{claim}[theorem]{Claim}
\newcommand{\true}{\mathsf{True}\xspace}
\newcommand{\True}{\textsf{True}\xspace}
\newcommand{\false}{\mathsf{False}\xspace}

\newenvironment{claimProof}[1]{\par\noindent\textbf{Proof of Claim}\space#1}{\hfill $\blacksquare$}

\newcommand{\sP}{\mathcal{P}\xspace}

\newcommand{\N}{\mathbb{N}\xspace}
\newcommand{\capacity}{\mathsf{channelCapacity}\xspace}

\usepackage{caption}

\SetKwInOut{Input}{input}
\SetKwInOut{Output}{output}

\newenvironment{proof}{\noindent \textbf{Proof.}}{\hfill$\blacksquare$}
\newenvironment{proofsketch}{\noindent\textbf{Proof Sketch.}}{\hfill$\Box$}

\newcommand{\Section}[1]{\section{#1}}
\newcommand{\Subsection}[1]{\subsection{#1}}
\newcommand{\Subsubsection}[1]{\noindent \textbf{#1.}~~~~~~}

\usepackage[utf8]{inputenc}

\begin{document}
	%
\title{Self-Stabilizing Indulgent Zero-degrading Binary Consensus}
	
	\author{Oskar Lundstr\"om~\thanks{Department of Computer Science and Engineering, Chalmers University of Technology, Gothenburg, SE-412 96, Sweden, E-mail: \texttt{osklunds@student.chalmers.se}.} \and Michel Raynal~\thanks{Institut Universitaire de France IRISA, ISTIC Universit\'{e} de Rennes, Rennes cedex, 35042, France, and Polytechnic University, Hong Kong. E-mail: \texttt{michel.raynal@irisa.fr}.} \and Elad M.\ Schiller~\thanks{Department of Computer Science and Engineering, Chalmers University of Technology, Gothenburg, SE-412 96, Sweden, E-mail: \texttt{elad@chalmers.se}.}}
	
%
%
%

\maketitle

\begin{abstract}
	
Guerraoui proposed an indulgent solution \ems{for the binary consensus problem. Namely,} he showed that an arbitrary behavior of the failure detector never violates safety requirements even if it compromises liveness. Consensus implementations are often used in a repeated manner. Dutta and Guerraoui proposed a zero-degrading solution, \ie during system runs in which the failure detector behaves perfectly, a node failure during one consensus instance has no impact on the performance of future instances. 
	
	Our study, which focuses on indulgent zero-degrading binary consensus, aims at the design of an even more robust communication abstraction. We do so through the lenses of \emph{self-stabilization}---a very strong notion of fault-tolerance. In addition to node and communication failures, self-stabilizing algorithms can recover after the occurrence of \emph{arbitrary transient faults}; these faults represent any violation of the assumptions according to which the system was designed to operate (as long as the algorithm code stays intact).  
	
	This work proposes the first, to the best of our knowledge, self-stabilizing algorithm for indulgent zero-degrading binary consensus for time-free message-passing systems prone to detectable process failures. The proposed algorithm has an $\bigO(1)$ stabilization time (in terms of asynchronous cycles) from arbitrary transient faults. \remove{Moreover, the communication costs of our algorithm are similar to the ones of the non-self-stabilizing solution by Guerraoui and Raynal. The main difference is that our proposal uses repeated communications until the consensus object is deactivated.} Since the proposed solution uses an $\Omega$ failure detector, we also present the first, to the best of our knowledge, self-stabilizing asynchronous $\Omega$ failure detector, which is a variation on the one by Most{\'{e}}faoui, Mourgaya, and Raynal. 
\end{abstract}


\Section{Introduction}
\label{sec:intro}
We propose a self-stabilizing implementation of \emph{binary consensus} objects for time-free (aka asynchronous) message-passing systems whose nodes may fail-stop. We also show a self-stabilizing asynchronous construction of eventual leader failure detector, $\Omega$.

\Subsection{Background and motivation} 
With the information revolution, everything became connected, \eg banking services, online reservations, e-commerce, IoTs, automated driving systems, to name a few. All of these applications are distributed, use message-passing systems, and require fault-tolerant implementations. Designing and verifying these systems is notoriously difficult since the system designers have to cope with their asynchronous nature and the presence of failures. The combined presence of failures and asynchrony creates uncertainties (from the perspective of individual processes) with respect to the application state. Indeed, Fischer, Lynch, and Paterson~\cite{DBLP:journals/jacm/FischerLP85} showed that, in the presence of at least one (undetectable) process crash, there is no deterministic algorithm for determining the state of an asynchronous message-passing system in a way that can be validly agreed on by all non-faulty processes.

This work is motivated by applications whose state is replicated over several processes in a way that emulates a finite-state machine. In order to maintain consistent replicas, each process has to apply the same sequence of state-transitions according to different sources of (user) input. To this end, one can divide the problem into two: (i) propagate the user input to all replicas, and (ii) let each replica perform the same sequence of state-transitions. The former challenge can be rather simply addressed via uniform reliable broadcast~\cite{DBLP:books/sp/Raynal18,hadzilacos1994modular}, whereas the latter one is often considered to be at the problem core since it requires all processes to agree on a common value, \ie the order in which all replicas apply their state transitions. In other words, the input must be totally ordered before delivering it to the emulated automaton. 

It was observed that the agreement problem of item (ii) can be generalized. Namely, the consensus problem requires each process to propose a value, and all non-faulty processes to agree on a single decision, which must be one of the proposed values. The problem of fault-tolerant consensus was studied extensively in the context of time-free message passing-systems. The goal of our work is to broaden the set of failures that such solutions can tolerate.

\begin{figure}
	\begin{center}
				\BBB\BBB
		\includegraphics[scale=0.4, clip]{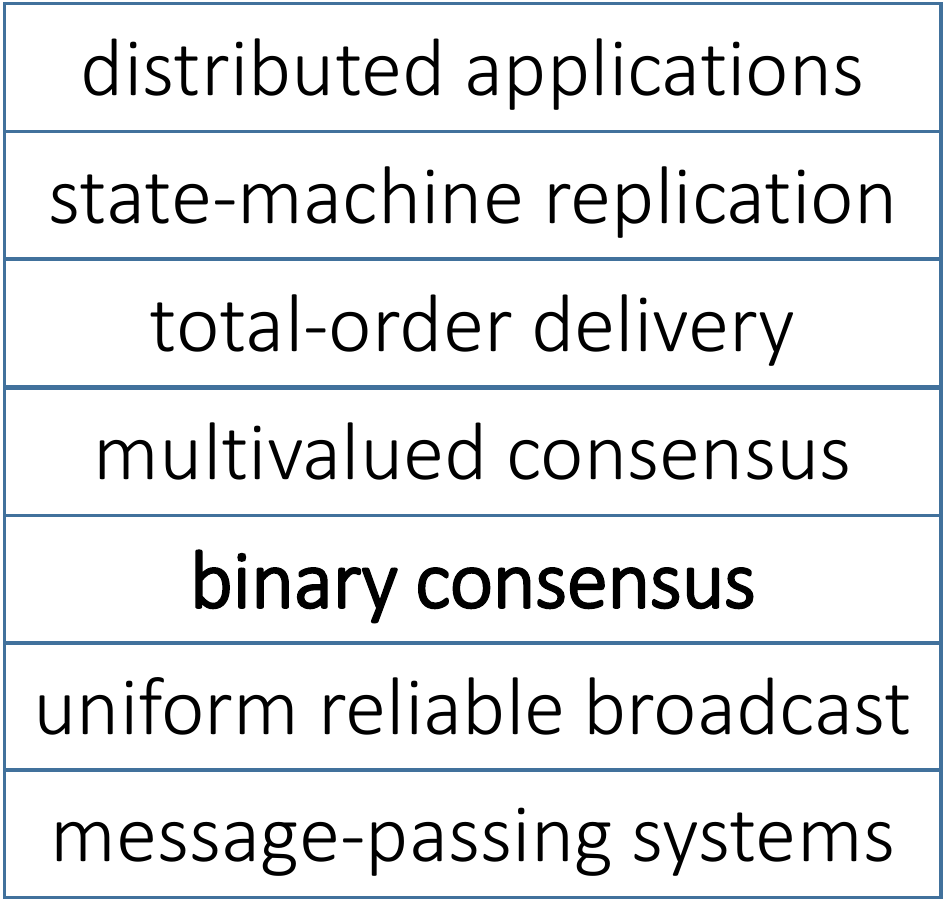}
	\end{center}
	\BBB\B
	\caption{\label{fig:suit}{The studied problem \reduce{of binary consensus} (in bold) in the context of a relevant protocol suite}.}
		\BBB\BB
\end{figure}

\Subsection{Problem definition and scope} 
Definition~\ref{def:consensus} states the consensus problem. When the set, $V$, of values that can be proposed, includes just two values, the problem is called binary consensus. Otherwise, it is called multivalued consensus. Existing solutions for multivalued consensus often use binary consensus algorithms. Figure~\ref{fig:suit} depicts the relation to other problems in the area, which were mentioned earlier.    

\begin{definition}[The consensus problem]
	\label{def:consensus}
	Every process $p_i$ has to propose a value $v_i \in V$ via an invocation of the $\mathsf{propose}_i(v_i)$ operation, where $V$ is a finite set of values. Let $\mathit{Alg}$ be an algorithm that solves consensus. $\mathit{Alg}$ has to satisfy \emph{safety} (\ie validity, integrity, and agreement) and \emph{liveness} (\ie termination).
	\begin{itemize}
		\item \textbf{Validity.} Suppose that $v$ is decided. Then, $\mathsf{propose}(v)$ was invoked by some process.
		\item \textbf{Integrity.} Suppose a process decides. It does so at most once.
		\item \textbf{Agreement.} No two processes decide different values.
		\item \textbf{Termination.} All non-faulty processes decide.
	\end{itemize}
\end{definition}

As mentioned earlier, consensus cannot be solved in asynchronous message-passing systems that are prone to failures, as weak as even the crash of a single process~\cite{DBLP:journals/jacm/FischerLP85}. Unreliable failure detectors~\cite{DBLP:journals/jacm/ChandraT96} are often used to circumvent such impossibilities. For a given failure detector class, Guerraoui~\cite{DBLP:conf/podc/Guerraoui00} proposed an indulgent solution, namely, he showed that an arbitrary behavior of the failure detector never violates safety requirements even if it compromises liveness. Consensus implementations are often used in a repeated manner. Dutta and Guerraoui~\cite{DBLP:conf/edcc/DuttaG02} proposed a zero-degrading solution, \ie during system runs in which the failure detector behaves perfectly, a failure during one consensus instance has no impact on the performance of future instances. We study solutions for indulgent zero-degrading binary consensus.

\Subsection{Fault Model} 
We study a time-free message-passing system that has no guarantees on the communication delay and the algorithm cannot explicitly access the local clock. Our fault model includes $(i)$ detectable fail-stop failures of processes, and $(ii)$ communication failures, such as packet omission, duplication, and reordering. 

In addition to the failures captured in our model, we also aim to recover from \emph{arbitrary transient faults}, \ie any temporary violation of assumptions according to which the system and network were designed to operate, \eg the corruption of control variables, such as the program counter, packet payload, and indices, \eg sequence numbers, which are responsible for the correct operation of the studied system, as well as operational assumptions, such as that at least a majority of nodes never fail. Since the occurrence of these failures can be arbitrarily combined, we assume that these transient faults can alter the system state in unpredictable ways. In particular, when modeling the system, we assume that these violations bring the system to an arbitrary state from which a \emph{self-stabilizing algorithm} should recover the system.

\Subsection{Related Work} 
The celebrated Paxos algorithm~\cite{DBLP:journals/tocs/Lamport98} circumvents the impossibility by Fischer, Lynch, and Paterson~\cite{DBLP:journals/jacm/FischerLP85} by assuming that failed computers can be detected by unreliable failure detectors~\cite{DBLP:journals/jacm/ChandraT96}. These detectors can eventually notify the algorithm about the set of computers that were recently up and connected. However, there is no bound on the time that it takes the algorithm to receive a correct version of this notification. It is worth mentioning that Paxos has inspired many veins of research, \eg~\cite[and references therein]{DBLP:journals/csur/RenesseA15}. We, however, follow the family of abstractions by Raynal~\cite{DBLP:books/sp/Raynal18} due to its clear presentation that is easy to grasp as well as the fact that it can facilitate efficient implementations.


\Subsubsection{Non-self-stabilizing solutions} 
The $\Omega$ class includes eventual leader failure detectors. Chandra, Hadzilacos, and Toueg~\cite{DBLP:journals/jacm/ChandraHT96} defined this class and showed that it is the weakest for solving consensus in asynchronous message-passing systems while assuming that at most a minority of the nodes may fail. In this work we study the $\Omega$ failure detector by Most{\'{e}}faoui, Mourgaya, and Raynal~\cite{DBLP:conf/dsn/MostefaouiMR03}. We note the existence of a computationally equivalent $\Omega$ failure detector by Aguilera \etal~\cite{DBLP:conf/podc/AguileraDFT04}, which explicitly accesses timers. Our study focuses on~\cite{DBLP:conf/dsn/MostefaouiMR03} since it is asynchronous.   


Guerraoui~\cite{DBLP:conf/podc/Guerraoui00} presented the design criterion of indulgence. Guerraoui and Lynch~\cite{DBLP:conf/sss/GuerraouiL06} studied this criterion formally. Raynal~\cite{DBLP:journals/tc/GuerraouiR04,DBLP:journals/cj/GuerraouiR07} generalized it and designed indulgent $\Omega$-based consensus algorithms.	
Dutta and Guerraoui~\cite{DBLP:conf/edcc/DuttaG02} introduced the zero-degradation criterion. The studied algorithm is by Guerraoui and Raynal~\cite{DBLP:journals/tc/GuerraouiR04} who presented an indulgent zero-degrading consensus algorithm for message-passing systems in which the majority of the nodes never fail, and $\Omega$-failure detectors are available. We have selected this algorithm due to its clear presentation and the fact that it matches the ``two rounds'' lower bound by Keidar and Rajsbaum~\cite{DBLP:journals/ipl/KeidarR03}. 
Hurfin \etal~\cite{DBLP:journals/tc/HurfinMR02} showed that zero-degradation can be combined with the versatile use of a family of failure detector for improving the efficiency of round-based consensus algorithms. Wu \etal~\cite{DBLP:journals/jpdc/WuCYR08} presented the notion of round-zero-degradation, which extend zero-degradation, and the notation of look-head. They presented algorithms that extend the ones by Hurfin \etal and can reduce the number of required rounds. We note that such extensions are also plausible for our solutions.  

\Subsubsection{Self-stabilizing solutions} 
We follow the design criteria of self-stabilization, which Dijkstra~\cite{DBLP:journals/cacm/Dijkstra74} proposed. A detailed pretension of self-stabilization was provided by Dolev~\cite{DBLP:books/mit/Dolev2000} and Altisen \etal~\cite{DBLP:series/synthesis/2019Altisen}. 

Blanchard \etal~\cite{DBLP:conf/netys/BlanchardDBD14} have a self-stabilizing failure detector for partially synchronous systems. They mention the class P of perfect failure detectors. Indeed, there is a self-stabilizing asynchronous failure detector for class P by Beauquier and Kekkonen{-}Moneta~\cite{DBLP:journals/ijsysc/BeauquierK97} and a self-stabilizing synchronous $\Omega$ failure detector by Delporte{-}Gallet, Devismes, and Fauconnier~\cite{DBLP:conf/sss/Delporte-GalletDF07}. We present the first, to the best of our knowledge, asynchronous $\Omega$ failure detector. Hutle and Widder~\cite{DBLP:conf/sss/HutleW05} present an impossibility result that connects fault detection, self-stabilization, and time-freedom as well as link capacity and local memory bounds. They explain how randomization can circumvent this impossibility for eventually perfect failure detector~\cite{DBLP:conf/pdcn/HutleW05}. Biely \etal~\cite{DBLP:conf/sss/BielyHPW07} connect between classes of deterministic failure detectors, self-stabilization, and synchrony assumptions. We follow the assumption made by Most{\'{e}}faoui, Mourgaya, and Raynal~\cite{DBLP:conf/dsn/MostefaouiMR03} regarding communication patterns, which is another way to circumvent such impossibilities.      

The consensus problem was not extensively studied in the context of self-stabilization. The notable exceptions are by Dolev \etal~\cite{DBLP:journals/jcss/DolevKS10} and Blanchard \etal~\cite{DBLP:conf/netys/BlanchardDBD14}, which presented the first practically-self-stabilizing solutions for share-memory and message-passing systems, respectively. 
%
%
We note that practically-self-stabilizing systems, as defined by Alon \etal~\cite{DBLP:journals/jcss/AlonADDPT15}\reduce{ and clarified by Salem and Schiller~\cite{DBLP:conf/netys/SalemS18}}, do not satisfy Dijkstra's requirements, \ie practically-self-stabilizing systems do not guarantee recovery within a finite time after the occurrence of transient faults. Moreover, the message size of Blanchard \etal is polynomial in the number of processes, whereas ours is a constant (that depends on the number of bits it takes to represent a process identifier). The origin of the design criteria of practically-self-stabilizing systems can be traced back to Dolev \etal~\cite{DBLP:journals/jcss/DolevKS10}, who provided a practically-self-stabilizing solution for the consensus problem in shared memory systems, whereas we study message-passing systems. \reduce{It is worth mentioning that the work of Blanchard \etal has lead to the work of Dolev \etal~\cite{DBLP:journals/jcss/DolevGMS18}, which considers a practically-self-stabilizing emulation of state-machine replication, which has the same task of the state-machine replication in Figure~\ref{fig:suit}. However, Dolev \etal's solution is based on virtual synchrony by Birman and Joseph~\cite{DBLP:journals/tocs/BirmanJ87}, where the one in Figure~\ref{fig:suit} consider censuses. We also note that earlier self-stabilizing algorithms for state-machine replications were based on group communication systems and assumed execution fairness~\cite{DBLP:journals/tpds/DolevS03,DBLP:journals/acta/DolevS04,DBLP:journals/tmc/DolevSW06}.}

There are other self-stabilizing algorithms that are the result of transformations of non-self-stabilizing yet solutions, such as for atomic snapshots~\cite{DBLP:conf/netys/GeorgiouLS19}, uniform reliable broadcast~\cite{selfStabURB}, set-constraint delivery broadcast~\cite{DBLP:conf/icdcs/ICDCS2020} and coded atomic storage~\cite{DBLP:journals/corr/abs-1806-03498}.

\Subsection{Our contribution} 
We present a fundamental module for dependable distributed systems: a self-stabilizing algorithm for indulgent zero-degrading binary consensus for time-free message-passing systems that are prone to detectable node fail-stop failures. 

The design criteria of indulgence and zero-degradation are essential for facilitating efficient distributed replication systems and self-stabilization is imperative for significantly advancing the fault-tolerance degree of future replication systems. \ems{Indulgence means that the safety properties, \eg agreement, are never compromised even if the underlying model assumptions are never satisfied. Zero-degrading means that the process failures that occurred before the algorithm starts have no impact on its efficiency, which depends only on the failure pattern that occur during the system run.} To the best of our knowledge, we are the first to provide a solution for binary consensus that is indulgent, zero-degrading and can tolerate a fault model as broad as ours. Our model includes detectable fail-stop failures, communication failures, such as packet omission, duplication, and reordering as well as arbitrary transient faults. The latter can model any \ems{temporary} violation of the assumptions according to which the system was designed to operate (as long as the algorithm code stays intact). 

In the absence of transient faults, our solution achieves consensus within an optimal number of communication rounds (without assuming fair execution). After the occurrence of any finite number of arbitrary transient faults, the system recovers within an asymptotically optimal time (while assuming fair execution). Namely, the stabilization time is in $\bigO(1)$ (in terms of asynchronous cycles). As in Guerraoui and Raynal~\cite{DBLP:journals/tc/GuerraouiR04}, each node uses a bounded amount of memory. Moreover, the communication costs of our algorithm are similar to the non-self-stabilizing one by Guerraoui and Raynal~\cite{DBLP:journals/tc/GuerraouiR04}. The main difference is in the period after a node has decided. Then, it has to broadcast the decided value. At that time, the non-self-stabilizing solution in~\cite{DBLP:journals/tc/GuerraouiR04} terminates whereas our self-stabilizing solution repeats the broadcast until the consensus object is deactivated by the invoking algorithm. This is along the lines of a well-known impossibility~\cite[Chapter 2.3]{DBLP:books/mit/Dolev2000} \ems{stating} that self-stabilizing systems cannot terminate. Also, it is easy to trade the broadcast repetition rate with the speed of recovery from transient faults.  

We also propose the first, to the best of our knowledge, self-stabilizing asynchronous $\Omega$ failure detector, which is a variation on Most{\'{e}}faoui, Mourgaya, and Raynal~\cite{DBLP:conf/dsn/MostefaouiMR03}. We show transient fault recovery within the time it takes all non-crashed processes to exchange messages among themselves. The use of local memory and communication costs are asymptotically the same as the one of~\cite{DBLP:conf/dsn/MostefaouiMR03}. The key difference is that we deal with the ``counting to infinity'' scenario, which transient fault can introduce. The proposed self-stabilizing solution uses \ems{a} trade-off parameter, $\delta$, that can balance between the solution's vulnerability (to elect a crashed node as a leader even in the absence of transient faults) and the time it takes to elect a non-faulty leader (after the occurrence of the last transient fault). Note that $\delta \in \mathbb{Z}^+$ can be a predefined constant.

As an extension, we also discuss how to transform the (non-self-stabilizing) randomized algorithm for binary consensus by Ben{-}Or~\cite{DBLP:conf/podc/Ben-Or83} to a self-stabilizing one (Section~\ref{sec:disc}).

\reduce{
\Subsection{Organization} We state our system settings in Section~\ref{sec:sys}. Section~\ref{sec:omega} present our self-stabilizing asynchronous $\Omega$ failure detector. Section~\ref{sec:back} includes a brief overview of the earlier algorithm that has led to the proposed solution. Our self-stabilizing algorithm is proposed in Section~\ref{sec:binary}; it considers unbounded counters. \ems{The correctness proof appears in Section~\ref{sec:correctness}.} 
%
%
We sketch an extension and conclude in Section~\ref{sec:disc}.
}

%

\smallskip

\Section{System settings}
\label{sec:sys}
We consider a time-free message-passing system that has no guarantees on the communication delay. Moreover, there is no notion of global (or universal) clocks and the algorithm cannot explicitly access the local clock (or timeout mechanisms). The system consists of a set, $\sP$, of $n$ fail-prone nodes (or processors) with unique identifiers. Any pair of nodes $p_i,p_j \in \sP$ have access to a bidirectional communication channel, $\mathit{channel}_{j,i}$, that, at any time, has at most $\capacity \in \N$ packets on transit from $p_j$ to $p_i$ (this assumption is due to a well-known impossibility~\cite[Chapter 3.2]{DBLP:books/mit/Dolev2000}). 

%
%
In the \emph{interleaving model}~\cite{DBLP:books/mit/Dolev2000}, the node's program is a sequence of \emph{(atomic) steps}. Each step starts with an internal computation and finishes with a single communication operation, \ie a message $send$ or $receive$. The \emph{state}, $s_i$, of node $p_i \in \sP$ includes all of $p_i$'s variables and $\mathit{channel}_{j,i}$. The term \emph{system state} (or configuration) refers to the tuple $c = (s_1, s_2, \cdots,  s_n)$. We define an \emph{execution (or run)} $R={c[0],a[0],c[1],a[1],\ldots}$ as an alternating sequence of system states $c[x]$ and steps $a[x]$, such that each $c[x+1]$, except for the starting one, $c[0]$, is obtained from $c[x]$ by $a[x]$'s execution. 

\Subsection{Task specification}
\label{sec:spec}
The set of \emph{legal executions} ($LE$) refers to all the executions in which the requirements of the task $T$ hold. In this work, $T_{\text{binCon}}$ denotes the task of binary consensus, which Definition~\ref{def:consensus} specifies, and $LE_{\text{binCon}}$ denotes the set of executions in which the system fulfills $T_{\text{binCon}}$'s requirements. Definition~\ref{def:consensus} considers the \ems{$\mathsf{propose}(s,k,v)$ operation. We refine the definition of $\mathsf{propose}(s,k,v)$ to include the value of $s$ and $k$ that we describe next.} Moreover, we specify how the decided value is retrieved. We clarify that it can be either via the returned value of the $\mathsf{propose}()$ operation (as in the studied algorithm~\cite{DBLP:journals/tc/GuerraouiR04}) or via the returned value of the $\done(s,k)$ operation (as in the proposed solution). The proposed solution is tailored for the protocol suite presented in Figure~\ref{fig:suit}. Thus, we consider multivalued consensus objects that use an array, $BC[]$, of $n$ binary consensus objects, such as the one by~\cite[Chapter 17]{DBLP:books/sp/Raynal18}, where $n=|\sP|$ is the number of nodes in the system. Moreover, we organize these multivalued consensus objects in an array, $CS[]$, of $M$ elements, where $M \in \mathbb{Z}^+$ is a predefined constant. We note that in case the algorithm that uses $CS[]$ runs out of consensus objects, a global restart procedure can be invoked, such as the one in~\cite{DBLP:conf/netys/GeorgiouLS19}, Section~5. Thus, it is possible to have bounded sequence numbers for multivalued objects.

\Subsection{The fault model and self-stabilization}
A failure occurrence is a step that the environment takes rather than the algorithm.

\Subsubsection{Benign failures}
\label{sec:benignFailures}
When the occurrence of a failure cannot cause the system execution to lose legality, \ie to leave $LE$, we refer to that failure as a benign one. 
%
%
The studied consensus algorithms are prone to \emph{fail-stop failures}, in which nodes stop taking steps. We assume that at most $t<|P|/2$ node may fail and that unreliable failure detectors~\cite{DBLP:journals/jacm/ChandraT96} can detect these failures. The studied failure detector constructions consider (undetectable) \emph{crash failures}.
%
%
%
We consider solutions that are oriented towards time-free message-passing systems and thus they are oblivious to the time in which the packets arrive and depart. We assume that any message can reside in a communication channel only for a finite period. Also, the communication channels are prone to packet failures, such as omission, duplication, reordering. However, if $p_i$ sends a message infinitely often to $p_j$, node $p_j$ receives that message infinitely often. We refer to the latter as the \emph{fair communication} assumption.

\Subsubsection{Arbitrary transient faults}
\label{sec:arbitraryTransientFaults}
We consider any temporary violation of the assumptions according to which the system was designed to operate. We refer to these violations and deviations as \emph{arbitrary transient faults} and assume that they can corrupt the system state arbitrarily (while keeping the program code intact). The occurrence of an arbitrary transient fault is rare. Thus, our model assumes that the last arbitrary transient fault occurs before the system execution starts~\cite{DBLP:books/mit/Dolev2000}. Also, it leaves the system to start in an arbitrary state.

\Subsubsection{Dijkstra's self-stabilization criterion}
\label{sec:Dijkstra}
An algorithm is \textit{self-stabilizing} with respect to the task of $LE$, when every (unbounded) execution $R$ of the algorithm reaches within a finite period a suffix $R_{legal} \in LE$ that is legal. Namely, Dijkstra~\cite{DBLP:journals/cacm/Dijkstra74} requires $\forall R:\exists R': R=R' \circ R_{legal} \land R_{legal} \in LE \land |R'| \in \mathbb{Z}^+$, where the operator $\circ$ denotes that $R=R' \circ R''$ concatenates $R'$ with $R''$.

%
%

\Subsubsection{Complexity Measures}
\label{sec:timeComplexity}
The complexity measure of self-stabilizing systems, called \emph{stabilization time}, is the time it takes the system to recover after the occurrence of the last transient fault. Next, we provide the assumptions needed for defining this period.  

We \emph{do not assume} execution fairness in the absence of transient faults. We say that a system execution is \emph{fair} when every step that is applicable infinitely often is executed infinitely often and fair communication is kept. After the occurrence of the last transient fault, we assume the system execution is \emph{temporarily} fair until the system reaches a legal execution, as in Georgiou \etal~\cite{DBLP:conf/netys/GeorgiouLS19}. 

Since asynchronous systems do not consider the notion of time, we use the term (asynchronous) cycles as an alternative way to measure the period between two system states in a fair execution. The first (asynchronous) cycle (with round-trips) of a fair execution $R=R' \circ R''$ is the shortest prefix $R'$ of $R$, such that each non-failing node executes at least one complete iteration (of the do forever loop) in $R'$. The second cycle in $R$ is the first cycle in $R''$, and so on. 
We clarify the term complete iteration (of the do forever loop). It is well-known that self-stabilizing algorithms cannot terminate their execution and stop sending messages~\cite[Chapter 2.3]{DBLP:books/mit/Dolev2000}. Moreover, their code includes a do forever loop. Let $N_i$ be the set of nodes with whom $p_i$ completes a message round trip infinitely often in $R$. Suppose that immediately after the state $c_{begin}$, node $p_i$ takes a step that includes the execution of the first line of the do forever loop, and immediately after system state $c_{end}$, it holds that: (i) $p_i$ has completed the iteration of $c_{begin}$ and (ii) every request message $m$ (and its reply) that $p_i$ has sent to any non-failing node $p_j \in \sP$ during the iteration (of the do forever loop) has completed its round trip. In this case, we say that $p_i$'s complete iteration starts at $c_{begin}$ and ends at $c_{end}$.


\Subsection{Uniform reliable broadcast}
We assume the availability of a self-stabilizing uniform reliable broadcast (URB)~\cite{selfStabURB}, which requires that if a node (faulty or not) delivers a message, then all non-failing nodes also deliver this message~\cite{hadzilacos1994modular}. The task specifications consider an operation for URB broadcasting of message $m$ and an event of URB delivery of message $m$. The requirements include URB-validity, \ie there is no spontaneous creation or alteration of URB messages, URB-integrity, \ie there is no duplication of URB messages, as well as URB-termination, \ie if the broadcasting node is non-faulty, or if at least one receiver URB-delivers a message, then all non-failing nodes URB-deliver that message. Note that the URB-termination property considers both faulty and non-faulty receivers. This is the reason why this type of reliable broadcast is named \emph{uniform}. This work also assumes that the operation for URB broadcasting message $m$ returns a transmission descriptor, $\txD$, which is the unique message identifier. Moreover, the predicate $\exist(\txD)$ holds whenever the sender knows that all non-failing nodes in the system have delivered $m$. The implementation of $\exist(\txD)$ can just test that the local buffer does not include any record with the message identifier $\txD$. The solution in~\cite{selfStabURB} can facilitate the implementation of $\exist()$ since the self-stabilizing algorithm in~\cite{selfStabURB} removes obsolete records of messages that were delivered by all non-faulty receivers.  

\Subsection{Unreliable failure detectors}
\label{sec:ext}
Chandra and Toueg~\cite{DBLP:journals/jacm/ChandraT96} introduced the concepts of failure patterns and unreliable failure detectors. Chandra, Hadzilacos, and Toueg~\cite{DBLP:journals/jacm/ChandraHT96} proposed the class $\Omega$ of eventual leader failure detectors. It is known to be the weakest failure detector class to solve consensus. A pedagogical presentation of these failure detectors is given in Raynal~\cite{DBLP:books/sp/Raynal18}. 

\Subsubsection{Failure patterns}
Any execution $R:=(c[0],a[0],c[1],a[1],\ldots)$ can have any number of failures during its run. $R$'s failure pattern is a function $F:\mathbb{Z}^+ \rightarrow 2^\sP$, where $\mathbb{Z}^+$ refers to an index of a system state in $R$, which in some sense represents (progress over) time, and $2^\sP$ is the power-set of $\sP$, which represents the set of failing nodes in a given system state. $F(\tau)$ denotes the set of failing nodes in system state $c_{\tau}\in R$. Since we consider fail-stop failures, $F(\tau) \subseteq F(\tau + 1)$ holds for any $\tau \in \mathbb{Z}^+$. Denote by $\mathit{Faulty}(F)\subseteq \sP$ the set of nodes that eventually fail-stop in the (unbounded) execution $R$, which has the failure pattern $F$. Moreover, $\mathit{Correct}(F)=\sP \setminus \mathit{Faulty}(F)$. For brevity, we sometimes notate these sets as $\mathit{Correct}$ and $\mathit{Faulty}$. 

\Subsubsection{Eventual leader failure detectors}
\label{sec:Omega}
This class allows $p_i \in \sP$ to access a read-only local variable $leader_i$, such that $\{leader_i\}_{1\leq i \leq n}$ satisfy the $\Omega$-validity and $\Omega$-eventual leadership requirements, where $leader^\tau_i$ denotes $leader_i$'s value in system state $c_\tau \in R$ of system execution $R$. $\Omega$-validity requires that $\forall i: \forall \tau : leader^\tau_i$ contains a node identity. $\Omega$-eventual leadership requires that $\exists \ell \in  \mathit{Correct}(F), \exists c_\tau \in R: \forall \tau' \geq \tau : \forall i \in  \mathit{Correct}(F): leader^{\tau'}_i= \ell$. These requirements imply that a unique and non-faulty leader is eventually elected, however, they do not specify when this occurs and how many leaders might co-exist during an arbitrarily long (yet finite) anarchy period. \ems{Moreover, no processor can detect the ending of this period of anarchy.}


\Section{Failure Detectors for the $\Omega$  Class}
\label{sec:omega}
We study a non-self-stabilizing construction of an $\Omega$ failure detector (Section~\ref{sec:Omega}) and propose its self-stabilizing variant.


%

\Subsection{Non-self-stabilizing $\Omega$  failure detector}
Algorithm~\ref{alg:somegaNon} presents the non-self-stabilizing $\Omega$  failure detector by Most{\'{e}}faoui, Mourgaya, and Raynal~\cite{DBLP:conf/dsn/MostefaouiMR03}; the boxed code lines are irrelevant to~\cite{DBLP:conf/dsn/MostefaouiMR03} since we use them to present our self-stabilizing solution. \ems{Note that, in addition to the assumptions described in Section~\ref{sec:spec}, Most{\'{e}}faoui, Mourgaya, and Raynal make the following operational assumptions (Section~\ref{sec:opetionalAssumptions}), which are asynchronous by nature.}   

\Subsection{Operational assumptions}
\label{sec:opetionalAssumptions}
Algorithm~\ref{alg:somegaNon} follows Assumption~\ref{def:EMP}. Let us observe Algorithm~\ref{alg:somegaNon}'s communication pattern of queries and responses. Node $p_i$ broadcasts $\mathsf{ALIVE()}$ queries repeatedly until the arrival of the corresponding $\mathsf{RESPONSE}()$ messages from $(n-t)$ receivers (the maximum number of messages from distinct nodes it can wait for without risking being blocked forever). For the sake of a simple presentation (and without loss of generality), it is assumed that nodes always receive their own responses. We refer to the first $(n-t)$ replies to a query that $p_i$ receives as the winning responses. The others are referred to as the losing since, after a crash, the failing nodes cannot reply.

\begin{assumption}[Eventual Message Pattern]
	\label{def:EMP}
	In any execution $R$, there is a system state $c_{\tau} \in R$, a non-faulty $p_i \in \sP$, and a set $Q$ of $(t+1)$ nodes, such that, after $c_{\tau}$, each node $p_j \in Q$ always receives a winning response from $p_i$ to each of its queries (until $p_j$ possibly crashes). (Note that the time until the system reaches $c_{\tau}$, the \ems{identity of} $p_i$ and the set $Q$ need not be explicitly known by the nodes.)
\end{assumption}

\Subsection{Variables}
The local state includes $r_i$, which is initialized to $0$ and is used for indexing $p_i$'s current round of alive queries and responses. Moreover, the array $count[]$ counts the number of suspicions, \eg $count_i[j]$ counts from zero the number of times $p_i$ suspected $p_j$. Also, the $recFrom$ set, which is initialized to $\sP$, has the identities of the nodes which responded to the most recent alive query. When the application layer accesses the variable $leader$, Algorithm~\ref{alg:somegaNon} returns the identity of the least suspected node (line~\ref{ln:sleaderRead}).

\Subsection{Algorithm description}
Algorithm~\ref{alg:somegaNon} repeatedly executes a do forever loop (lines~\ref{ln:startLoop} to~\ref{ln:srecFromRESPONSE}), which broadcasts $\mathsf{ALIVE}(r,\bullet)$ messages (line~\ref{ln:ssendIAMALIVE}) and collects their replies, which are the $\mathsf{RESPONSE}(\mathit{rJ},\bullet)$ messages (lines~\ref{ln:sRESPONSErecived} and~\ref{ln:ssendRESPONSE}). In this message exchange, every $p_i \in \sP$ uses a round number, $r_i$, to facilitate asynchronous rounds without any coordination linking the rounds of different nodes. Moreover, there is no limit on the number of steps any node takes to complete an asynchronous round.

\Subsubsection{The do forever loop}
Each iteration of the do forever loop includes actions (1) to (3).
	(1) Node $p_i$ broadcasts $\mathsf{ALIVE}(r_i,count_i)$ queries (line~\ref{ln:ssendIAMALIVE}), and waits for $(n-t)$ replies, \ie $\mathsf{RESPONSE}(\mathit{rJ}, \mathit{recFromJ})$ messages from $p_j \in \sP$ (line~\ref{ln:ssendRESPONSE}), where $r_i$ and $\mathit{rJ}$ are matching round numbers. Moreover, $count_i$ is an array in which, as said before, $count_i[k]$ stores the number of times $p_i$ suspected $p_k \in \sP$. Also, $\mathit{recFromJ}$ is a set of the identities of the responders to $p_j$'s most recent query (lines~\ref{ln:srecFromRESPONSE} and~\ref{ln:ssendRESPONSE}).
%
	(2) By aggregating into $prevRecFrom_i$ all the arriving $\mathit{recFromJ}$ sets (line~\ref{ln:sRESPONSErecived}), $p_i$ can estimate that any $p_j: j\notin prevRecFrom_i$ that does not appear in any of these sets is faulty. Thus, $p_i$ increment $count_i[j]$ (line~\ref{ln:scountPlusOne}).
	(3) The iteration of the do forever loop ends with a local update to $p_i$'s $recFrom_i$ (line~\ref{ln:srecFromRESPONSE}).

\Subsubsection{Processing of arriving queries}
Upon $\mathsf{ALIVE}(\mathit{rJ}, \mathit{countJ})$ arrival from $p_j$, node $p_i$ merges the arriving data with its own (line~\ref{ln:smaxCountK}), and replies with $\mathsf{RESPONSE}(\mathit{rJ}, recFrom_i)$ (line~\ref{ln:ssendRESPONSE}). This reply includes $p_j$'s round number, $\mathit{rJ}$, which is not linked to $p_i$'s round number, $r_i$.


\begin{algorithm*}[t!]
	\begin{\algSize}
		
		\smallskip
		
		\noindent \textbf{local constant, variables and their initialization:} (Initialization is optional in the context of self-stabilization.)\\
		\fbox{\textbf{const} $\delta$} \tcc*{max gap between the extrema of count values}
		
		$r:=0$ \tcc*{current round number}
		$recFrom:= \sP$ \tcc*{set of identities of the processors that replied to the most recent query} 
		$count[0..n\text{-}1]:=[0,\ldots,0]$ \tcc*{the number of times each processor was suspected}
		
		\smallskip
		
		\textbf{operation} $\mathsf{leader}()$ \label{ln:sleader} \{\textbf{let} $(\bull,x) := \min\{(count[k], k)\}_{p_k \in \sP}$; \Return($x$)\}\label{ln:sleaderRead}
		
		\smallskip
		
		\fbox{\textbf{macro} $\counts():=\{count[k]:p_k \in \sP\}$}\; 
		
		\smallskip
		
		\fbox{\textbf{macro} $\checkC():=$\textbf{ if }{$\max \counts() - \min \counts()\ems{>} \delta$\label{ln:consistencyAssertion}}\textbf{ then foreach } {$p_k \in \sP$} \textbf{do}} \fbox{$count[k]\gets \max \{count[k], (\max \counts() -\delta)\}$;}
		
		\smallskip
		
		\textbf{do forever} \label{ln:startLoop}\Begin{			
			\Repeat{$\mathsf{RESPONSE}(\mathit{rJ},$ \fbox{$\bull,$} $\mathit{recFromJ})$ \textbf{\emph{received from}} $(n-t)$ {\emph{processors}}\label{ln:sRESPONSErecived}}{\lForEach{$j \neq i$}{\textbf{send} $\mathsf{ALIVE}(r, count)$ \textbf{to} $p_j$\label{ln:ssendIAMALIVE}}}
			
			\textbf{let} $prevRecFrom := \cup$ sets of $\mathit{recFromJ}$ received in line~\ref{ln:sRESPONSErecived}\;
			\lForEach{$j \notin prevRecFrom$\fbox{$: count[j] < \delta+\min \counts()$}}{$count[j] \gets count[j]+1$\label{ln:scountPlusOne}}
			$recFrom \gets \{$processors from which $p_i$ has received $\mathsf{RESPONSE}(\mathit{rJ},\bullet)$ in line~\ref{ln:sRESPONSErecived}$\}$\label{ln:srecFromRESPONSE}\;
			\fbox{$\checkC()$}\label{ln:checkLoop}\; 
		}
		
		\smallskip
		
		\textbf{upon} $\mathsf{ALIVE}(\mathit{rJ}, \mathit{countJ})$ \textbf{arrival from} $p_j$ \label{ln:suponIAMALIVE}\Begin{
			\lForEach{$p_k \in \sP$}{$count[k] \gets \max(count[k], \mathit{countJ}[k])$\label{ln:smaxCountK}}
			\fbox{$\checkC()$}\label{ln:checkAlive}\; 
			\textbf{send} $\mathsf{RESPONSE}(\mathit{rJ}, \fbox{count,} \mathit{recFrom})$ to $p_j$\label{ln:ssendRESPONSE}
		}	
		
		\smallskip
		
		\fbox{\textbf{upon} $\mathsf{RESPONSE}(\mathit{rJ}, \mathit{countJ}, \mathit{recFromJ})$ \textbf{arrival from} $p_j$} \label{ln:suponRESPONSE}\Begin{
			\fbox{\lForEach{$p_k \in \sP$}{$count[k] \gets \max(count[k], \mathit{countJ}[k])$\label{ln:smaxCountK2}}}
			
			\fbox{$\checkC()$}\label{ln:checkResponse}\; 
		}
		
		\smallskip
		
		\caption{\label{alg:somegaNon}An $\Omega$ construction; code for $p_i$. (Only the self-stabilizing version includes the boxed code lines.)}
	\end{\algSize}
\end{algorithm*}

\Subsection{Self-stabilizing $\Omega$  failure detector}
When including the boxed code lines, Algorithm~\ref{alg:somegaNon} presents an unbounded self-stabilizing variation of the $\Omega$  failure detector in~\cite{DBLP:conf/dsn/MostefaouiMR03}. (As mentioned before, Section~5 in~\cite{DBLP:conf/netys/GeorgiouLS19} \ems{explains} how to convert such unbounded self-stabilizing algorithms to bounded ones.) Note that in~\cite{DBLP:conf/dsn/MostefaouiMR03}, all non-crashed nodes converge to a constant value that is known to all correct nodes whereas the counters of crashed nodes increase forever, see claim C2 and C3 of Theorem~97 in~\cite{DBLP:books/sp/Raynal18}. Thus, the proposed algorithm includes \ems{the following} differences from~\cite{DBLP:conf/dsn/MostefaouiMR03}.

Algorithm~\ref{alg:somegaNon} makes sure that any non-failing node does not ``hide'' a value that is too high in $count_i[x]$ without sharing it with all correct nodes. In the context of self-stabilization, such a value can appear due to a transient fault. To that end, Algorithm~\ref{alg:somegaNon} includes the field $count$ in the $\mathsf{RESPONSE}()$ message (line~\ref{ln:ssendRESPONSE}) so that the receiver can merge the arriving data with the local one (line~\ref{ln:smaxCountK2}). 

Algorithm~\ref{alg:somegaNon} also avoids ``counting to infinity'' since, in the context of self-stabilization, a transient fault can set the counters \ems{to arbitrary values. 
For example, suppose that the counter values that non-faulty nodes associates with all crashed nodes is zero. Also suppose that the counters associated with any non-faulty node is extremely high, say, $M=2^{62}$.
We must not require the system to count from zero to $M$ before it is guaranteed that a non-crashed leader is elected, because it would take more than 146 years to do (of we assume the speed of one nanosecond per communication round).}
Thus, the proposed solution limits the difference between the extrema counter values in any local array to be \ems{less} than $\delta$, where $\delta$ is a predefined constant. One can view $\delta$ as a trade-off parameter between the solution vulnerability (to elect a crashed node as a leader even in the absence of transient faults) and the time it takes to elect a non-faulty leader (after the occurrence of the last transient fault and after the system has reached $c_\tau$ that satisfies the eventual message pattern assumption, cf. Assumption~\ref{def:EMP}). \ems{\Ie on the one hand, if the value of $\delta$ is set too low, processors that sporadically slow down might be elected, while on the other hand, for very large values of $\delta$, say, $M$, the time it takes to recover after the occurrence of the last transient faults can be extremely long.}

\Subsubsection{Correctness} 
Definitions~\ref{def:safeConfig} and~\ref{def:complete} are needed for showing that Algorithm~\ref{alg:somegaNon} brings the system to a legal execution (Theorem~\ref{thm:consistencyAssertion}).   

\begin{definition}[Algorithm~\ref{alg:somegaNon}'s consistent system state]
	\label{def:safeConfig}
	Suppose that $\max \counts_i() - \min \counts_i()\leq \delta$ holds in $c \in R$ for any non-faulty $p_i \in\sP$.  
	In this case, we say $c$ is consistent.
\end{definition}

\begin{definition}[Complete execution of Algorithm~\ref{alg:somegaNon}]
	\label{def:complete}
	Let $R$ be an execution of Algorithm~\ref{alg:somegaNon}. Let $c, c'' \in R$ denote the starting system states of $R$, and respectively, $R''$, for some suffix $R''$ of $R$. We say that message $m$ is \emph{completely delivered} in $c$ if the communication channels do not include $\mathsf{ALIVE}(r,\bullet)$ nor $\mathsf{RESPONSE}(r,\bullet)$ messages. 
	Suppose that $R=R' \circ R''$ has a suffix $R''$, such that for any $\mathsf{ALIVE}(r,\bullet)$ or $\mathsf{RESPONSE}(r,\bullet)$ message $m$ that is not completely delivered in $c''$, it holds that $m$ does not appear in $c$.
	In this case, we say that $R''$ is complete with respect to $R$.
\end{definition}

\BB \begin{theorem}[Convergence]
	\label{thm:consistencyAssertion}
	(i) Once every non-failing processor completes at least one iteration of the do forever loop (lines~\ref{ln:startLoop} to~\ref{ln:checkLoop}) or receive at least one message (lines~\ref{ln:suponIAMALIVE} or~\ref{ln:suponRESPONSE}), the system reaches a consistent state. (ii) Every infinite execution $R=R' \circ R''$ of Algorithm~\ref{alg:somegaNon} reaches within a finite number of steps suffix $R''$, such that $R''$ is complete with respect to $R$ (Definition~\ref{def:complete}).       
\end{theorem}

\BBB \begin{proof}
	Lines~\ref{ln:checkLoop},~\ref{ln:checkAlive}, and~\ref{ln:checkResponse} imply invariant (i). Invariant (ii) is implied by the assumption that any message can reside in a communication channel only for a finite time (Section~\ref{sec:benignFailures}). 
\end{proof}
 
 
\BB \begin{theorem}[Closure]
	\label{thm:fdClosed}
	Let $R$ be an execution of Algorithm~\ref{alg:somegaNon} that starts in a consistent system state. Suppose that $R$ has an eventual message pattern \ems{(Assumption~\ref{def:EMP}).}
	Algorithm~\ref{alg:somegaNon} demonstrates in $R$ a construction of the eventual leader failure detector, $\Omega$.
\end{theorem}

\remove{\
	
	BB \begin{proofsketch}
	In the context of Algorithm~\ref{alg:somegaNon}, we say that $p_i \in \sP$ inhibits the increment of $count_i[x]$ in line~\ref{ln:scountPlusOne} when $x \notin prevRecFrom$ holds but $count_i[x] < \delta+\min \counts_i()$ does not. Suppose that, for a given $p_x \in \sP$, there is $p_k \in \sP$ that, during $R$, either increments $count_k[x]$  in line~\ref{ln:scountPlusOne} or inhibits such increments for a bounded number of times. In this case, we say that $count_k[x]$ is bounded. 
	%
	%
	The proof shows that all non-crashed nodes converge to a bounded value that becomes known to all correct nodes whereas the counters of crashed nodes increase forever. One can obtain the proof by taking Theorem~97 in~\cite{DBLP:books/sp/Raynal18} and substantiating `increasing $count_i[x]$' in Claim 2 with `the increment of $count_i[x]$ or its inhibition'. 
\end{proofsketch}

} 


	\BBB \begin{proof}
		In the context of Algorithm~\ref{alg:somegaNon}, we say that $p_i \in \sP$ inhibits the increment of $count_i[x]$ in line~\ref{ln:scountPlusOne} when $x \notin prevRecFrom$ holds but $count_i[x] < \delta+\min \counts_i()$ does not. Suppose that, for a given $p_x \in \sP$, there is $p_k \in \sP$ that, during $R$, either increments $count_k[x]$  in line~\ref{ln:scountPlusOne} or in inhibits such increments for a bounded number of times. In this case, we say that $count_k[x]$ is bounded. In all other cases, we say that $count_k[x]$ is unbounded. Given a failure pattern $F()$, we define: $PL = \{x : \exists i \in \mathit{Correct}(F) : count_i[x]$ is bounded$\}$, and $\forall i \in \mathit{Correct}(F) : PL_i = \{x : count_i[x]$ is bounded$\}$, where the set of processor identities, $PL$, stands for ``potential leaders''. These definitions imply $\forall i \in \mathit{Correct}(F) : PL_i \subseteq PL$.
		
		The rest of the proof shows that correct processors share identical sets of potential leaders ($PL$), which non-empty (Lemmas~\ref{thm:PLemptyset}), and include only correct processors (Lemmas~\ref{thm:PLsubseteqCorrectF} and~\ref{thm:iCorrectFplIpl}). The proof ends by showing that the processors in $PL$ can only be suspected, \ie their counters are incremented (or inhibited from being incremented), a bounded number of times, and this number is eventually the same at each non-faulty processor (Lemma~\ref{thm:twoNonFaultyCountIK}). Thus, all correct processors eventually elect the processor that was suspected for the smallest number of times.
		
		\begin{lemma}
			\label{thm:PLemptyset}
			$PL\neq \emptyset$
		\end{lemma}
		\BBB \begin{proof}
			Since Assumption~\ref{def:EMP} holds, there mus be a system state $c_{\tau_0} \in R$, a processor $p_i$ and a set $Q$ of $(t+1)$ processors for which at any state after $c_{\tau_0}$, any non-failing processor $p_j \in Q$ receives winning responses from $p_i$ for any of $p_i$'s queries. Due to the assumptions that $|Q|>t$ and that there are at most $t$ faulty processors, $Q$ includes at least one non-faulty processor. Let $\tau\geq\tau_0$ be a time after which no more processors fail.
			
			Processor $p_k \in \sP: k \in \mathit{Correct}(F)$ does not stop sending its query (line~\ref{ln:ssendIAMALIVE}) until it receives $\mathsf{RESPONSE}()$ messages from $(n-t)$ processors. Moreover, after $c_\tau$, at least $(t+1)$ processors get winning responses from $p_i$. Therefore, the system eventually reaches a state $c_{\tau_k} \in R: \tau \leq \tau_k$ after which $i \in prevRecFrom_k$ holds (line~\ref{ln:ssendRESPONSE}. Thus, $p_k$ stops incrementing (or inhibiting the increment) of $count_k[i]$ at line~\ref{ln:scountPlusOne}.
			
			Since $p_k$ is any correct processors, the system eventually reaches the state $c_{\max \{\tau_x\}_{x\in \mathit{Correct}(F)}} \in R$, it holds that $\forall x,y \in \mathit{Correct}(F):count_x[i] = count_y[i] = M_i \in \mathbb{Z}^+$. In other words, due to the repeated exchange of messages between any pair of non-faulty processors, these processors has a constant value for $count[i]$.
		\end{proof}
		
		\begin{lemma}
			\label{thm:PLsubseteqCorrectF}
			$PL \subseteq \mathit{Correct}(F)$.
		\end{lemma}
		\BBB \begin{proof}
			We show that for every $x \notin \mathit{Correct}(F)$, it holds that $p_i:i\in \mathit{Correct}(F)$ increments (or inhibits the increment) of $count_i[x]$ for an unbounded number of times during $R$. The rest of the proof is implied by the fact that non-faulty processors never stop exchanging messages among themselves and merge the arriving information upon message arrival (lines~\ref{ln:smaxCountK} and~\ref{ln:smaxCountK2}).
			
			Suppose that all the faulty processors have crashed (and their messages $\mathsf{RESPONSE}()$ have been received) before $c_\tau \in R$. Let $p_i$ and $p_j$ be non-faulty processors, and $p_x$ a faulty one. We observe invariants (i) to (iv), which imply the proof.
			(i) Since $p_x$ cannot respond to any of $p_j$'s queries, it holds that $x \notin \mathit{rF}_j$, where $x \notin \mathit{rF}_j$ is the value of $recFrom_j$ (which is assigned in line~\ref{ln:srecFromRESPONSE}) in any system state, $c'_\tau$, that appears in $R$ after $c_\tau$. 
			(ii) Due to invariant (i), it holds that $x \notin \mathit{pRF}_j$, where $x \notin \mathit{pRF}_j$ is the value of $prevRecFrom_i$ (which is assigned in line~\ref{ln:sRESPONSErecived}) in any system state, $c''_\tau$, that appears in $R$ after $c'_\tau$.
			(iii) Due to invariant (ii), after $c''_\tau$, every execution of line~\ref{ln:countPlusOne} implies an increment of $count_i[x]$ (or the inhibition of an increment).    
			(iv) Since $p_i$ sends an unbounded number of queries, invariant (iii) implies that $count_i[x]$ is incremented (or the inhibited from incrementing) for an unbounded number of times during $R$. 
		\end{proof}
		
		\begin{lemma}
			\label{thm:iCorrectFplIpl}
			$(i \in \mathit{Correct}(F)) \Rightarrow (PL_i = PL)$
		\end{lemma}
		\BBB \begin{proof}
			Recall that $PL_i \subseteq PL$ (by the definitions of $PL$ and $PL_i$). Therefore, $PL \subseteq PL_i$ implies the proof and $PL_i \subseteq Correct(F)$ (due to Lemma~\ref{thm:PLsubseteqCorrectF}). 
			Let assume that $k \in PL$ and show that $k \in PL_i$. That is, we assume that there are $k, j\in \mathit{Correct}(F)$ for which the constant $M_k$ is the highest value stored in $count_j[k]$ throughout $R$. In order to prove that $k \in PL_i$, we show that $count_i[k]$ is also bounded. Since $count_j[k] \leq M_k$ throughout $R$, the repeated exchange of $\mathsf{ALIVE}()$ and $\mathsf{RESPONSE}()$ messages between the correct processors $p_i$ and $p_j$ (line~\ref{ln:ssendIAMALIVE}, lines~\ref{ln:suponIAMALIVE} to~\ref{ln:smaxCountK} and lines~\ref{ln:suponRESPONSE} to~\ref{ln:smaxCountK2}), implies that $count_i[k]\leq M_k$ throughout $R$ (due to the fact that $M_k$ is a constant).	
		\end{proof}	
		
		\begin{lemma}
			\label{thm:twoNonFaultyCountIK}
			Let $i, j\in \mathit{Correct}(F)$. Suppose that $R$ has a suffix $R''$ during which $count_i[k]=M_k$ always hold, where $M_k$ is a constant. Then, $count_j[k]=M_k$ also holds throughout $R''$.
		\end{lemma}
		\BBB \begin{proof}
			This is due to the repeated exchange of $\mathsf{ALIVE}()$  and $\mathsf{RESPONSE}()$ messages between $p_i$ and $p_j$  (line~\ref{ln:ssendIAMALIVE}, lines~\ref{ln:suponIAMALIVE} to~\ref{ln:smaxCountK} and lines~\ref{ln:suponRESPONSE} to~\ref{ln:smaxCountK2}).
		\end{proof}
	\end{proof}
	This ends the proof of Theorem~\ref{thm:fdClosed}.

\Section{Background: Non-self-stabilizing Zero-degrading Binary Consensus}
\label{sec:back}
\begin{algorithm*}[t!]
	\begin{\algSize}
		
		\smallskip
		
		\noindent \textbf{local variables and their initialization:}\\
		$r:=0$ \tcc*{current round number}
		$est[0..1]:=[\bot,\bot]$ \tcc*{local decision estimates at the beginning of phases 0 and 1}
		
		\smallskip
		
		\textbf{operation} $\mathsf{propose}(v)$ \label{ln:scdBroadcast} \Begin{
			
			$(est[], r) \gets ([v,\bot],0)$	\tcc*{$\bot$ denotes a default value that cannot be proposed} 
			\While{$\true$}{{
					$r \gets r +1$\;
					\tcc{Phase 0: select a value with the help of  $\Omega$}
					\textbf{let} $myLeader := \mathsf{leader}$\label{ln:letMyLeaderNon} \tcc*{read $\Omega$}
					\Repeat{$[\mathrm{PHASE}(0,r ,\bullet)$ \emph{received from} $n-t$ \emph{nodes}$] \land [\mathrm{PHASE}(0,r,\bullet)$ \emph{received from} $p_{myLeader} \lor myLeader \neq \mathsf{leader}]$\label{ln:until0Non}}{\textbf{broadcast\xspace} $\mathrm{PHASE}(0,r, est[0],myLeader)$\label{ln:broadcast0Non}}
					
					\If{$[\mathrm{PHASE}(0,r,\bullet,\ell)$ \emph{received from more than} $n/2$ \emph{nodes}$] \land [\mathrm{PHASE}(0,r, v,\bullet)$ \emph{received from} $p_{\ell}]$\label{ln:result0Non}}{$est[1] \gets v$ \textbf{else} $est[1] \gets \bot$}
					

					\tcc{Here, we have $((est_i[1] \neq \bot) \land (est_j[1] \neq \bot)) \implies (est_i[1] = est_j[1] = v)$}
					\tcc{Phase 1: try to decide on an $est[1]$ value}
					
					\lRepeat{$[\mathrm{PHASE}(1,r ,\bullet)$ \emph{received from} $n-t$ \emph{nodes}$]$}{\textbf{broadcast\xspace} $\mathrm{PHASE}(1,r, est[1])$}
					
					
					\Switch{$\{rec: \mathrm{PHASE}(1,r, rec)$ {has been received}$\}$\label{ln:letREcEst2Non}}{
						\lCase{$\{v\}$}{\{\textbf{broadcast} $\mathrm{DECIDE}(v)$; $\Return(v)$\}\label{ln:vBroadcastNon}}
						\lCase{$\{v,\bot\}$}{$est[0] \gets v$\label{ln:vBotEstNon}}
						\lCase{$\{\bot\}$}{\textbf{continue}\label{ln:continueNon}}
					}
					
					%
			}}
		}
		
		\smallskip
		
		\textbf{upon} $\mathsf{DECIDE}(v)$ arrival from $p_j$ \textbf{do} \{\textbf{broadcast} $\mathrm{DECIDE}(v)$; $\Return(v)$;\label{ln:finalDeci}\}
		
		\smallskip
		
		\caption{\label{alg:consensusBinaryNon}Guerraoui-Raynal~\cite{DBLP:journals/tc/GuerraouiR04}'s non-self-stabilizing indulgent zero-degrading binary consensus; code for $p_i$}
	\end{\algSize}
\end{algorithm*}

%

Algorithm~\ref{alg:consensusBinaryNon} is a non-self-stabilizing $\Omega$-based binary consensus algorithm that is indulging and zero-degrading. \ems{For the sake of a simpler presentation of the correctness proofs, Algorithm~\ref{alg:consensusBinaryNon}'s line enumeration continues the one of Algorithm~\ref{alg:somegaNon}.}


\Subsection{Algorithm structure}
Algorithm~\ref{alg:consensusBinaryNon} proceeds in asynchronous rounds that combine, each, two phases. The algorithm aims to have, by the end of phase zero, the same value, which is named the estimated value. This selection is done by a leader, \ems{whose} election is facilitated by the $\Omega$ failure detector. Next, during phase one, the algorithm tests the success of phase zero. The challenge here is that, due to the asynchronous nature of the system, not all nodes run the same round simultaneously. Therefore, the test considers the agreement on the round number, the leader identity, and the proposed value. Moreover, just before deciding on any value, say $v$, the deciding node broadcasts a $\mathrm{DECIDE}(v)$ message. Upon $\mathrm{DECIDE}(v)$ arrival, the receiver repeats the broadcast of the arriving \ems{message} before deciding. Algorithm~\ref{alg:consensusBinaryNon} executes the ``decide'' action by returning with $v$ from $\mathsf{propose}(v)$'s invocation. \ems{This technique of `broadcast repetition' basically lets Algorithm~\ref{alg:consensusBinaryNon} to invoke a reliable broadcast of the decided value. }    

\Subsubsection{The system behavior during phase zero}
The objective of phase zero of round $r$ is to let all nodes to store in $est[1]$ the same value. Once that happens, a decision can be taken during phase one of round $r$. As we are about to explain, that objective is guaranteed to be achieved once a single leader is elected.

The main challenge that phase zero addresses is the provision of the safety property, \ie no two different decisions are made, during $\Omega$'s anarchy period in which there is no single non-faulty elected leader. To that end, phase zero makes sure that the \emph{quasi-agreement} property  always holds before anyone enters phase one of round $r$, where $((est_i[1] \neq \bot) \land (est_j[1] \neq \bot)) \implies (est_i[1] = est_j[1] = v)$ is the property definition. This means that, if $est_i[1]  = v \neq \bot$ holds, from the perspective of $p_i$, it can decide $v$. Moreover, if $est_i[1]  = \bot$ holds, then from the perspective of $p_i$, it is not ready to decide any value. Therefore, a system state that satisfies the quasi-agreement property allows the individual nodes to decide during phase one on the same value (when $est_i[1] = est_j[1] = v$), or defer the decision to the next round (when $est_i[1] = \bot$). In order to satisfy the quasi-agreement property by the end of phase zero, each $p_i \in \sP$ performs actions (1) and (2), which imply Corollary~\ref{thm:quasiAgreement}.

\begin{corollary}
	\label{thm:quasiAgreement}
	The \emph{quasi-agreement} property holds immediately before $p_i \in \sP$ enters phase one of any round. 
\end{corollary}


Action (1): \reduce{Processor} $p_i$ stores in  $myLeader_i$ the value of $leader_i$ \ems{(line~\ref{ln:letMyLeaderNon}),} which is the interface to the $\Omega$ failure detector, before broadcasting the message $\mathrm{PHASE}(0,r, est_i[0],myLeader_i)$ (line~\ref{ln:broadcast0Non}). It then waits until it hears from $n-t$ nodes on the same round (line~\ref{ln:until0Non}). Since there are at most $t$ crashed nodes, waiting for more than $n-t$ nodes jeopardizes the system liveness. Moreover, $t < n/2$ and thus any set of $n-t$ nodes is a majority set, which contains at least one correct node. Processor $p_i$ may stop broadcasting also when it receives a $\mathrm{PHASE}(0,r,\bullet)$ message from its leader, \ie $p_{\mathit{myLeader}_i}$, or when a new leader is elected, \ie $\mathit{myLeader}_i \neq leader_i$.
	
Action (2): after the above broadcast, $p_i$'s assignment to $est_i[1]$ (line~\ref{ln:result0Non}) satisfies the quasi-agreement property  by making sure that (i) a majority of nodes consider $p_\ell$ as their leader when they broadcast the $\mathrm{PHASE}(0,r,\bullet, \ell)$, and (ii) $p_i$ received $\mathrm{PHASE}(1,r,v,\bullet)$ from $p_\ell$. In other words, if (i) and (ii) hold, $p_i$ can assign $v$ to $est_i[1]$, which is $p_\ell$'s value in $est_\ell[0]$ at the start of round $r$. Otherwise, $est_i[1]$ gets $\bot$. Due to the majority intersection property, no two majority sets can have two different unique leaders. Therefore, it cannot be that \ems{$est1_i[r] = v \neq \bot$ and $est1_j[r] = v' \neq \bot$} with $v \neq v'$.

\reduce{Corollary~\ref{thm:quasiAgreement} is implied by the above.}

\Subsubsection{The system behavior during phase one}
During this phase, $p_i$ broadcasts $\mathrm{PHASE}(1,r, est_i[1])$ until it hears from $n-t$ nodes. By the quasi-agreement property, $\exists v \in V: \forall p_j \in \sP : est_j[1] = \bot \lor est_j[1] = v \neq \bot$ holds during round $r$. Thus, for the set of all received estimated values, $rec_i\in \{ \{v\}, \{v, \bot\}, \{\bot\} \}$ (line~\ref{ln:letREcEst2Non}) holds. For the $rec_i = {v}$ case, $p_i$ can broadcast $\mathrm{DECIDE}(v)$ before deciding $v$ (line~\ref{ln:vBroadcastNon}). For the $rec_i = \{v, \bot\}$ case, $p_i$ uses $v$ during round $r+1$ as the new estimated value $est_i[0]$ since some other node might have decided $v$ (line~\ref{ln:vBotEstNon}). For the $rec_i=\{\bot\}$ case, $p_i$ continues to round $r+1$ without modifying $est_i[0]$ (line~\ref{ln:continueNon}). Note that, at any round $r$, it cannot be the case that both $rec_i = \{v\}$ and $rec_j = \{\bot\}$ hold, since \ems{$p_i$'s} broadcast of $\mathrm{DECIDE}(v)$ implies that it had received $\mathrm{PHASE}(1,r_i,v)$ from a majority of nodes. Due to the majority intersection property, there is at least one $\mathrm{PHASE}(1,r_i,v)$ arrival to any $p_j \in \sP$ that executes line~\ref{ln:letREcEst2Non} since it also received $\mathrm{PHASE}(1,r_i,\bullet)$ \reduce{messages} from a majority. Thus, $rec_j = \{\bot\}$ cannot hold. 

\Subsubsection{The necessity of broadcasting $v$ before deciding on it}
Algorithm~\ref{alg:consensusBinaryNon} has to take into consideration the case in which not all nodes decide during round $r$. \Eg a majority of nodes might decide on round $r$, while a minority of them continues to round $r+1$ during which it must not wait in vain to hear from a majority. By broadcasting $\mathrm{DECIDE}(v)$ before deciding $v$, Algorithm~\ref{alg:consensusBinaryNon} allows the system to avoid such bad situations since once $p_i$ decides, it is guaranteed that eventually, all correct nodes decide.

%
%

\begin{algorithm*}[t!]
	\begin{\algSizeSmall}
		
				\noindent \textbf{variables:}  
		$\mathit{seq}$ {is the sequence number of the multivalued consensus object}; $k$ {is the node index, $p_k \in \sP$};  $r:=0$ {is the current round number}; $est[0..2]:=[\bot,\bot,\bot]$ {are the local decision estimates at the beginning of phases 0 and 1 as well as the decided value at entry 2}; $\mathit{myLeader}$ {is the current identity of the leader}; $\mathit{newR}$ {is the round number aggregated from all received values}; and $\txD$ {is the URB transmission descriptor for sharing the decision};
		
		\smallskip
		
		\textbf{operation} $\mathsf{propose}(s,k,v)$ \label{ln:scdBroadcastM} \{\textbf{if} {$\reduce{\test(s) \land }CS[\ell] \neq \dCS \land CS[\ell].BC[\ell']=\dBS$}\label{ln:binProposeInvoke} \textbf{where $(\ell,\ell'):=(s \bmod M,k $ $\bmod n)$ then} {$CS[\ell].BC[\ell'].(\mathit{seq},k,r,est$, $\mathit{myLeader},\mathit{newR},\txD) \gets (s,k,0,[v,\bot,\bot],\mathsf{leader},0,\bot)$\}}
		
		
		\textbf{operation} $\done(s,k)$ \label{ln:doneBConsensus}\{\textbf{if} {$\reduce{\neg \test(s) \lor }CS[\ell]=\dCS\lor CS[\ell].BC[\ell']=\dBS$ \textbf{where} $(\ell,\ell'):=(s \bmod M,k \bmod n)$} \textbf{then} {\Return{$\bot$} \textbf{else} \Return{$(CS[\ell].BC[\ell'].est[2])$}}\}
		

		\textbf{operation} $\deactivate(s,k)$ \label{ln:deactivate}\{\textbf{if} {$\reduce{\neg \test(s) \lor }CS[\ell] \neq \dCS$ \textbf{then} $CS[\ell].BC[\ell'] \gets \bot$ \textbf{where} $(\ell, \ell') := (s \bmod M, k \bmod n)$}\}

		\smallskip
		
		\textbf{do forever} \{\ForEach{$(\ell, k) \in \{0,..., M\text{-}1 \} \times \{0,..., n\text{-}1 \} : CS[\ell] \neq \dCS \land x \neq \dBS$ $\land (x.\txD = \bot \lor \exist( x.\txD))$ $\mathbf{with }~ x's~ \mathbf{ \emph{fields}}$ $\mathit{seq},r,est[],\mathit{myLeader},\mathit{newR}$ \textbf{\emph{and}} $\txD$ \textbf{\emph{where}} $x := CS[\ell].BC[k]$\label{ln:doforeverBcon}}{{

				\lIf{$est[2] = \bot \land \txD \neq \bot \land \exist(\txD)$\label{ln:noactiveTranmisssion}}{$\txD \gets \bot$}
				
				\lIf{$est[2] \neq \bot \land (\txD = \bot \lor \exist(\txD))$\label{ln:txDurbReBroadcastIf}}{$\txD \gets \mathsf{urbBroadcast}~\mathrm{DECIDE}(\mathit{seq},k,v)$\label{ln:txDurbReBroadcast}; \textbf{continue}}
				
				$(r,myLeader) \gets (\max\{r,\mathit{newR}\}) +1,\mathsf{leader})$\label{ln:letMyLeader}\tcc{read $\Omega$}
				
				\tcc{Phase 0 : select a value with the help of $\Omega$}
				\Repeat{$(est[2] \neq \bot \lor \txD \neq \bot) \lor \{[\mathrm{PHASE}(0,\bull,\mathit{seq},k,r,\bullet)$ \emph{received from} $n-t$ \emph{nodes}$] \land [\mathrm{PHASE}(0,\bull,\mathit{seq},k,r,\bullet)$ \emph{received from} $p_{myLeader} \lor myLeader \neq \mathsf{leader}]\}$\label{ln:repeatPhase0End}}{\textbf{broadcast\xspace} $\mathrm{PHASE}(0,\true,\mathit{seq},k,r, est[0],myLeader,r)$\label{ln:repeatPhase0Start}}
				
				\lIf{$[\mathrm{PHASE}(0,\bull,\mathit{seq},k,r,\bull,\ell,\bull)$ \emph{received from a majority}$] \land [(0,\bull,\mathit{seq},k,r, v,\bullet)$ \emph{received from} $p_{\ell}]$}{$est[1] \gets v$ \textbf{else} $est[1] \gets \bot$\label{ln:est1GetsV}}
				
				\tcc{Phase 1 : try to decide on an $est[1]$ value}
				
				\Repeat{$(est[2] \neq \bot \lor \txD \neq \bot) \lor [\mathrm{PHASE}(1,\bull,\mathit{seq},k,r ,\bullet)$ \emph{received from} $n-t$ \emph{nodes}$]$\label{ln:repeatPhase1End}}{\textbf{broadcast\xspace} $\mathrm{PHASE}(1,\True,\mathit{seq},k,r, est[1], r)$\label{ln:repeatPhase1Start}}
				
				\textbf{let} $rec = \{est[1] : \mathrm{PHASE}(1,\bull,\mathit{seq},k,r, est[1])$ {was received}$\}$\label{ln:letREcEst2}\;
				
				\Switch{$rec$}{
					\lCase{$\{v\} \land \txD = \bot$}{$\txD \gets \mathsf{urbBroadcast}~\mathrm{DECIDE}(\mathit{seq},k,v)$\label{ln:txDurbBroadcast}}
					\lCase{$\{\bot,v\}$}{$est[0] \gets v$ \texttt{/* $\bot$ \ems{must not be} in $v$'s domain */}\label{ln:botVcase}}
					\lCase{$\{\bot\}$}{\textbf{continue}\label{ln:botOnlycase}}
				}
			}
		}


		\textbf{upon} $\mathsf{PHASE}(\mathit{nJ},\mathit{aJ},\mathit{sJ},\mathit{kJ},\mathit{rJ}, \mathit{vJ}, \mathit{myLeaderJ},\mathit{newRj})$ \textbf{arrival} \textbf{from} $p_j$ \textbf{do} \Begin{
			
			\lIf{$(\reduce{\neg \test(\mathit{sJ}) \lor }CS[\mathit{sJ} \bmod M] = \dCS) \land \mathit{aJ}$}{\{\textbf{send} $\mathrm{PHASE}(\mathit{nJ},\false,\mathit{sJ},\mathit{kJ},\mathit{rJ}, \mathit{vJ}, \mathsf{leader},\mathit{newR})$ \textbf{to} $p_j$; \Return\}\label{ln:phaseArrivalValidate}}
			
			\textbf{let} $O:=CS[\mathit{sJ} \bmod M].BC[\mathit{kJ} \bmod n]$\label{ln:letOR}\;
			
			\lIf{$O=\dBS$\label{ln:OdBS}}{$O.(\mathit{seq},r,est,\mathit{myLeader},\mathit{newR}) \gets (\mathit{sJ},\mathit{rJ},[\mathit{vJ},\bot,\bot],\mathsf{leader},\max\{\mathit{rJ},\mathit{newRj}\})$\label{ln:newRAgregateA}}
			\lElse{$O.\mathit{newR} \gets \max\{O.\mathit{rJ},O.\mathit{newR},\mathit{newRj}\}$\label{ln:newRAgregateB}}
			
			\lIf{$\mathit{nJ}=1 \land O.est[1] = \bot$}{$O.est[1] \gets \mathit{vJ}$\label{ln:nJ1OestGets}}
			
			\lIf{$\mathit{aJ}$}{$\mathbf{send}~\mathrm{PHASE}(\mathit{nJ},\false,\mathit{sJ},\mathit{kJ},\mathit{rJ}, \mathit{vJ}, O.\mathit{myLeader}$, $\max \{O.r,O.\mathit{newR}\})~ \mathbf{to}~ p_j$\label{ln:letORsend}}
		}
		
		
		\textbf{upon} $\mathsf{DECIDE}(\mathit{sJ},\mathit{kJ},\mathit{vJ})$ \textbf{arrival from} $p_j$ \textbf{do} \label{ln:decideArrival}\Begin{
			
			\If{$\reduce{\test(\mathit{sJ}) \land }CS[\mathit{sJ} \bmod M] \neq \dCS$\label{ln:CSsJMsCS}}{
				\textbf{let} $O:=CS[\mathit{sJ} \bmod M].BC[\mathit{kJ} \bmod n]$\label{ln:oAssigment}\;
				\lIf{$O=\dCS$\label{ln:initOthatisBot}}{$O.(\mathit{seq},r,est,\mathit{myLeader},\mathit{newR}) \gets (\mathit{sJ},0,[\mathit{vJ},\mathit{vJ},\bot],\mathsf{leader},0)$\label{ln:newRAgregateC}}
				\lIf{$O.est[2] = \bot$}{$O.est[2] \gets \mathit{vJ}$\label{ln:Oest2GetsV} \texttt{/* decide $v$ */}\label{ln:decideArrivalEnd}}
			}
		}
		
		
		\caption{\label{alg:consensusBinary}A self-stabilizing algorithm for indulgent zero-degrading binary consensus; code for $p_i$}
	\end{\algSizeSmall}
\end{algorithm*}

\Section{Self-stabilizing Indulgent Zero-degrading Binary Consensus}
\label{sec:binary}
Algorithm~\ref{alg:consensusBinary} is our self-stabilizing variation on Guerraoui and Raynal~\cite{DBLP:journals/tc/GuerraouiR04}. The main difference between the proposed solution and Algorithm~\ref{alg:consensusBinaryNon} occurs after a value was decided. Then, Algorithm~\ref{alg:consensusBinaryNon} broadcasts before terminating (lines~\ref{ln:vBroadcastNon} and~\ref{ln:finalDeci}) whereas our self-stabilizing solution repeats the broadcast until the consensus object is deactivated by the invoking algorithm. This follows a well-known impossibility~\cite[Chapter 2.3]{DBLP:books/mit/Dolev2000} that self-stabilizing systems cannot terminate. Specifically, in the context of self-stabilization, Algorithm~\ref{alg:consensusBinaryNon} can be started in a system state in which exactly half of the nodes are at the (normal) initial state of binary objects. Moreover, due to the presence of transient faults, the program counters of the other half of the nodes \ems{can} point to the return command in line~\ref{ln:vBroadcastNon}. Starting from this state will cause the system to violate the termination property. A self-stabilizing solution can avoid this violation by repeating the broadcast of the decided value until the consensus object is deactivated. Note that one can reduce the overhead of the proposed solution by simply lowering the broadcast repetition rate, which in turn extends the stabilization time.

%
%


\Subsection{Variables}
As explained in Section~\ref{sec:spec}, the proposed binary consensus objects are used by multivalued consensus objects, \ie $BC[]$ is an array of $n$ binary consensus objects and $CS[]$ is an array of $M$ multivalued consensus objects. 
%
%
The binary consensus objects of Algorithm~\ref{alg:consensusBinary} have the private variables, which store the sequence number of the multivalued consensus object, $\mathit{seq}$, a node index, $k:p_k \in \sP$, and current round number, $r$. Also, the results of phase $x \in \{0,1\}$ is stored $est[x]$ and $est[2]$ stores the decided value. Algorithm~\ref{alg:consensusBinary} also stores the current identity of the leader, $\mathit{myLeader}$, \ems{the} round number aggregated from all received values, $\mathit{newR}$, and the transmission descriptor of the reliable broadcast of the decided value $\txD$.
We say the binary object $CS[\ell].BC[k]$ is active when $CS[\ell]\neq \bot$ and $CS[\ell].BC[k]\neq \bot$. For a given active binary object $x:=CS[\ell].BC[k]$, we say that $x$ has an active reliable broadcast when $(x.\txD \neq \bot \land \exist(x.\txD))$, \ie $x.\txD$ stores a descriptor of a transmission that has not terminated. 

\Subsection{Message structure}
Algorithm~\ref{alg:consensusBinary} uses the $\mathsf{DECIDE}(\mathit{seq},\mathit{k},\mathit{v})$ and $\mathsf{PHASE}(\mathit{phase},\mathit{ackNeed},\mathit{seq},\mathit{k},\mathit{r}, \mathit{v}, \mathit{leader},\mathit{newR})$ messages, where the field $\mathit{phase}$ refers to the phase number, $\mathit{ackNeed}$ indicates whether a reply is needed, $\mathit{seq}$ is the sequence number, $\mathit{k}$ is the node index, $\mathit{r}$ is the round number, $\mathit{v}$ is the estimated value, $\mathit{leader}$ is the round leader, and $\mathit{newR}$ is the sender's round number. 

\Subsection{Interface operations}
The operation $\mathsf{propose}(s,k,v)$ \ems{(Section~\ref{sec:spec})} allows the invoking node to propose value $v$ with sequence number $s$ and node index $k$ (line~\ref{ln:scdBroadcastM}). The operation $\done(s,k)$ returns the decided value, if such decision occurred (line~\ref{ln:doneBConsensus}). Otherwise, $\bot$ is returned. 
The operation $\deactivate(s,k)$ assigns $\bot$ to $CS[s].BC[k]$ (line~\ref{ln:deactivate}).


\Subsection{The do forever loop (lines~\ref{ln:doforeverBcon} to~\ref{ln:txDurbBroadcast})} 
The nodes iterate over all active binary objects, $x$, that do not have an active reliable broadcast. In case $x$ has a decided value and it had an active transmission that has terminated (line~\ref{ln:noactiveTranmisssion}), $p_i$ initializes $x$'s transmission descriptor. Also, in case $x$ has a decided value but is has no active transmission (line~\ref{ln:txDurbReBroadcastIf}), $p_i$ broadcasts the decided value. In line~\ref{ln:letMyLeader}, $p_i$ increments the round number and sample the $\Omega$ failure detector. Algorithm~\ref{alg:consensusBinary} considers situations in which, due to a transient fault, the round numbers go out of sync. It does this by letting $\mathit{newR}$ aggregate the highest round number that is disseminated in each message exchange (lines~\ref{ln:newRAgregateA},~\ref{ln:newRAgregateB},~\ref{ln:letORsend}, and~\ref{ln:newRAgregateC}). Then, at the start of a new round, the highest known round number is used (line~\ref{ln:letMyLeader}).

\ems{Although the above example considers a case that can only happen before the start of the system execution, cf. Section~\ref{sec:arbitraryTransientFaults}, the system cannot know whether its current state is the starting one. Therefore, the system has to always be ready to recover from arbitrary transient faults. We also clarify that our model does not limit the number of nodes that can be affected by any arbitrary transient faults. It is only the example above that makes this assumption.}

\Subsubsection{Phase 0 (lines~\ref{ln:repeatPhase0Start} to~\ref{ln:est1GetsV})}
In this phase $p_i$ broadcasts $\mathrm{PHASE}(0,\true,s,k,r,$ $est[0],myLeader,r)$, such that the phase field is 0, acknowledgment is needed, the sequence number is $\mathit{seq}$, the node index is $k$, the round number is $r$, the estimated result is $est[0]$, the message leader is $myLeader$ and the message aggregated round number is $r$. This broadcasting repeats as long as the binary object neither has an active broadcast, nor stores a decided value. Moreover, the broadcasting continues until $\mathrm{PHASE}(0,\bull,\mathit{seq},k,r,\bullet)$ is received from $n-t$ nodes (which means that phase 0 messages were received from a majority of nodes during round $r_i$), or $\mathrm{PHASE}(0,\bull,\mathit{seq},k,r,\bullet)$ is received from $p_{myLeader} \lor myLeader \neq \mathsf{leader}]\}$ (which means that some nodes follow a leader different than $p_{myLeader_i}$ during $r_i$). Phase 0 ends by testing in line~\ref{ln:est1GetsV} whether a phase 0 message was received from a majority of nodes that have reported on the same leader, $p_\ell$, from which a message was received. If this is the case, $p_i$ uses the \ems{value,} $v$, received from $p_\ell$ as the estimated result for phase 1 by assigning $v$ to $est_i[1]$. Otherwise, $\bot$ is assigned.    

\Subsubsection{Phase 1 (lines~\ref{ln:repeatPhase1Start} to~\ref{ln:txDurbBroadcast})}
In this phase $p_i$ broadcasts $\mathrm{PHASE}(1,\True,\mathit{seq},k,r, est[1], r)$, such that the phase field is 1, acknowledgment is needed, the sequence number is $\mathit{seq}$, the node index is $k$, the round number is $r$, the estimated result is $est[1]$ and the message aggregated round number is $r$. As in phase 0, this broadcasting repeats as long as the binary object neither has an active broadcast, nor stores a decided value. Moreover, the broadcasting continues until $\mathrm{PHASE}(1,\bullet,\mathit{seq},k,r ,\bullet)$ was received from $n-t$ nodes (which means that phase 1 messages were received from a majority of nodes during round $r_i$). Phase 1 ends by testing the set, $rec_i$, of received estimated results during this phase (line~\ref{ln:letREcEst2}). 
By the quasi-agreement property (Section~\ref{sec:back}), $rec_i\in \{ \{v\}, \{v, \bot\}, \{\bot\} \}$ holds. When $rec_i = {v}$ holds, $p_i$ can reliably broadcast $\mathrm{DECIDE}(\mathit{seq},k,v)$ (line~\ref{ln:txDurbBroadcast}). When $rec_i = \{v, \bot\}$ holds, $p_i$ uses $v$ as the new estimated value $est_i[0]$ for round $r+1$ since some other node might have decided $v$ (line~\ref{ln:botVcase}). When $rec_i=\{\bot\}$ holds, $est_i[0]$ is unchanged before round $r+1$  (line~\ref{ln:botOnlycase}). 
%

\Subsection{The arrival of $\mathsf{PHASE}()$ messages}
This arrival updates (and even initializes) the local state of the binary consensus, $O_i$, that has the sequence number $\mathit{sJ}$ and node index $\mathit{kJ}$, where $\mathit{nJ},\mathit{aJ},\mathit{sJ},\mathit{kJ}$, $\mathit{rJ}, \mathit{vJ}, \mathit{myLeaderJ},\mathit{newRj}$ are the message fields. Before this, there is a need to test $\mathit{sJ}$ and validate that $CS[\mathit{sJ} \bmod M]$ is an active object. If this is not the case, a reply is sent to the sender (if $\mathit{aJ}$ indicates that this is needed) and the procedure returns (line~\ref{ln:phaseArrivalValidate}).

Line~\ref{ln:letOR} prepares the binary consensus object $O_i$ and line~\ref{ln:OdBS} tests whether $O_i$ needs to be initialized. Otherwise, Algorithm~\ref{alg:consensusBinary} updates the aggregated round number (line~\ref{ln:newRAgregateB}). Line~\ref{ln:nJ1OestGets} is applicable only for phase 1 messages. It tests whether $O_i$ has $\bot$ as its estimated result. When this is the case, $\mathit{vJ}$ is used as $O_i$'s estimated value. The procedure ends by acknowledging the sender, if needed (line~\ref{ln:letORsend}).

\Subsection{The arrival of $\mathsf{DECIDE}()$ message}
The arrival of $\mathsf{PHASE}(\mathit{sJ},\mathit{kJ},\mathit{vJ})$ messages can update (and even initialize) the decided value, $\mathit{vJ}$, of the binary consensus, $O_i$, that has the sequence number $\mathit{sJ}$ and node index $\mathit{kJ}$. Before this is done, there is a need to test $\mathit{sJ}$ and validate that $CS[\mathit{sJ} \bmod M]$ is an active object (line~\ref{ln:CSsJMsCS}). If this is the case, the procedure checks whether $O_i$ needs to be initialized together with the assignment of the decided value (line~\ref{ln:initOthatisBot}). Otherwise, line~\ref{ln:Oest2GetsV} simply assigns $\mathit{vJ}$ to $O_i.est[2]$.

\Section{Algorithm~\ref{alg:consensusBinary}'s Correctness}
\label{sec:correctness}
Theorems~\ref{thm:recoveryConsensusBinary} and~\ref{thm:clousureConsensusBinary} show the convergence and closure properties.

\BB \begin{theorem}[Algorithm~\ref{alg:consensusBinary}'s Convergence]
	\label{thm:recoveryConsensusBinary}
	Let $R$ be an execution of Algorithm~\ref{alg:consensusBinary}. Suppose that \reduce{for any sequence number $s$, any processor $p_i \in \sP$ and a step $a_i \in R$ in which $p_i$ calls $\test()$, it holds that $\test_i(s)=\true$. Also,} $\forall x \in \mathit{Correct}:CS_x[s \bmod M] \neq \bot$ holds in every system state of $R$. Moreover, $\exists p_j \in \sP : j \in \mathit{Correct}$ for which in every system state of $R$, it holds that $CS_i[s \bmod M] \neq \bot \land CS_i[s \bmod M].BC[k \bmod n] \neq \bot$ for some $k:p_k \in \sP$. Eventually the system reaches a state, $c \in R$, in which $\forall x \in \mathit{Correct} : CS_x[s \bmod M].BC[k \bmod n].est[2]\neq\bot$ holds. 
	%
\end{theorem}
\BBB \begin{proof}
	Claims~\ref{thm:testSmeanAllactive},~\ref{thm:termination}, and~\ref{thm:xDecides} imply the proof. 
	
	\BB \begin{claim}
		\label{thm:testSmeanAllactive}
		Suppose that $CS_i[s \bmod M].BC[k \bmod n].est[2]\neq\bot$ holds in $R$'s starting state. Eventually $\forall x \in \mathit{Correct}:CS_x[s \bmod M].BC[k \bmod n] \neq \bot$ holds throughout $R$.
	\end{claim}

	\BB \begin{claimProof}{\textbf{\ref{thm:testSmeanAllactive}.~~}}
		Every iteration of the do forever loop (lines~\ref{ln:doforeverBcon} to~\ref{ln:txDurbBroadcast}) includes the binary consensus object $CS_i[s \bmod M].BC[k \bmod n]$ (line~\ref{ln:doforeverBcon}). Note that $CS_i[s \bmod M].BC[k \bmod n].est[2]\neq\bot$ in the starting system state of $R$ implies that $CS_i[s \bmod M].BC[k \bmod n].est[2]\neq\bot$ holds throughout $R$ due to the theorem assumptions and the fact that Algorithm~\ref{alg:consensusBinary} never assigns $\bot$ to $CS_i[s \bmod M].BC[k \bmod n].est[2]$. Thus, the rest of the proof assumes, without loss of generality, that $CS_i[s \bmod M].BC[k \bmod n].est[2]\neq\bot$ holds throughout $R$.

		
		%
		Whenever $CS_i[s \bmod M].BC[k \bmod n].\txD \neq \bot$ holds, $\exist_i(\txD)$ holds eventually (\reduce{due to the}URB-termination\reduce{property}). Thus, the if-statement condition in line~\ref{ln:txDurbReBroadcastIf} holds eventually and $\mathrm{DECIDE}(s,k,v)$ is URB broadcast (line~\ref{ln:txDurbReBroadcast}). Recall that $v$'s domain does not include $\bot$ (line~\ref{ln:botVcase}). Upon the URB-delivery of $\mathrm{DECIDE}(s,k,v)$ at any $p_x: x \in \mathit{Correct}$, we have $\forall p_x \in \sP: x \in \mathit{Correct}:CS_x[s \bmod M].BC[k \bmod n] \neq \bot$ (lines~\ref{ln:decideArrival} to~\ref{ln:decideArrivalEnd}).      
		%
	\end{claimProof}
	
	\BB \begin{claim}
		\label{thm:termination}
		For any sequence number $s$, processor $p_k$'s proposal, and round $r>0$, there is no correct processor that considers indefinitely this consensus task at round $r$.
	\end{claim}
	\BB \begin{claimProof}{\textbf{\ref{thm:termination}.~~}}
		Without loss of generality, let us assume that $p_i \in  \sP$ does not decide during round $r$, \ie $CS_i[s \bmod M].BC[k \bmod n].est[2]=\bot$ throughout $R$. Generality is not lost due to the proof of Claim~\ref{thm:testSmeanAllactive} since the case in which $CS_i[s \bmod M].BC[k \bmod n].est[2]\neq\bot$ holds, implies that eventually $\forall p_x \in \sP: x \in \mathit{Correct}:CS_x[s \bmod M].BC[k \bmod n] \neq \bot$ holds, \ie termination.  
		%
		%
		%
		%
		Towards a contradiction, suppose that $r$ is the smallest round in which a correct processor $p_i$ executes indefinitely. The only two loops in which $p_i$ can continue to execute forever in round $r$ are the repeat-until loops in lines~\ref{ln:repeatPhase0Start} to~\ref{ln:repeatPhase0End} and~\ref{ln:repeatPhase1Start} to~\ref{ln:repeatPhase1End}. 
		
		By the choice of $r$ as well as lines~\ref{ln:letMyLeader},~\ref{ln:letOR}, and~\ref{ln:letORsend}, no correct processor can continue to execute forever in round $r'< r$. Therefore, $p_i$ receives $\mathrm{PHASE}(0,\bullet,s,k,r,\bullet)$ at least $(n-t)$ times. Moreover, if its current leader, $p_{myLeader_i}$, is correct, $p_i$ receives at least one $\mathrm{PHASE}(0,\bullet,s,k,r,\bullet)$ message from $p_{myLeader_i}$. Furthermore, if $p_{myLeader_i}$ is faulty, eventually it holds that $myLeader_i \neq leader_i$ (by $\Omega$'s eventual leadership). Thus, no correct processor $p_i$ can execute forever the repeat-until loop in lines~\ref{ln:repeatPhase0Start} to~\ref{ln:repeatPhase0End} during round $r$. By similar arguments, during phase one of round $r$, processor $p_i$ receives $\mathrm{PHASE}(1,\bullet,s,k,r,\bullet)$ messages at least $(n-t)$ times from the correct processors. Thus, during round $r$, processor $p_i$ does not execute forever the repeat-until loop in lines~\ref{ln:repeatPhase1Start} to~\ref{ln:repeatPhase1End}. Note that we have reached a contradiction with the assumption that $r$ is the smallest round in which a correct processor executes forever and therefore the claim is true.
	\end{claimProof}
	
	\BB \begin{claim}
		\label{thm:xDecides}
		Eventually only the correct nodes are alive and connected and $\forall x \in \mathit{Correct} : CS_x[s \bmod M].BC[k \bmod n].est[2]\neq\bot$. 
	\end{claim}
	\BB \begin{claimProof}{\textbf{\ref{thm:xDecides}.~~}}
		Assume, toward a contradiction, that no node ever decides with respect to sequence number $s$, $p_k \in \sP$, and proposal $v \in V$. Recall $\Omega$'s eventual leadership and the fact that faulty nodes eventually crash (by definition). Thus, Claim~\ref{thm:termination} implies the existence of a finite round number $r$ from which (a) only the correct nodes are alive and connected, as well as (b) all correct $p_i \in \sP$ share the same correct leader, \eg $p_d$, in $\mathit{myLeader}_i$.
		The end condition of the repeat-until loop in line~\ref{ln:repeatPhase0End} holds for $p_i$ eventually. This is because there are more than $n/2$ correct nodes. Each such node, including $p_x$, broadcasts $\mathrm{PHASE}(0, \bullet,s,k,r,v,\bullet)$ and receives at least $n-t$ times the messages $\mathrm{PHASE}(0, \bullet,s,k,r,\bullet)$ (cf. Claim~\ref{thm:termination}'s  proof). Once line~\ref{ln:repeatPhase0End}'s condition holds, by the same reasons, also the if-statement condition in line~\ref{ln:est1GetsV} holds as well. Thus, $p_i$ assigns $v$ to $CS[s].BC[k]_i.est[1]$, and during phase $1$ of round $r$, $p_i$ only sends $\mathrm{PHASE}(1, \bullet,s,k,r,v,\bullet)$. Since this is true for any correct $p_i$, it must be that $rec_i =\{v\}$ (line~\ref{ln:letREcEst2}). Therefore, every correct $p_i \in \sP$ executes line~\ref{ln:txDurbBroadcast}, in which $p_i$ URB-broadcasts $\mathrm{DECIDE}(s,k,v)$. Moreover, upon the URB delivery of $\mathrm{DECIDE}(s,k,v)$, every correct node decides in line~\ref{ln:Oest2GetsV} (Claim~\ref{thm:testSmeanAllactive}). 
		%
	\end{claimProof}
	
	This completes Theorem~\ref{thm:recoveryConsensusBinary}'s proof.
\end{proof}

Theorem~\ref{thm:clousureConsensusBinary} uses the definition of consistent executions.  
Let $p_i,p_k \in \sP$ be two nodes in the system and $s$ be a sequence number. Let $c$ be a system state in which the if-statement condition in line~\ref{ln:binProposeInvoke} holds with respect to $p_i$. Moreover, no communication channel include the messages $\mathsf{DECIDE}(sek=s,k=x,\bullet)$ and $\mathsf{PHASE}(\bullet,sek=s,k=x,\bullet)$. In this case, we say that $p_i$ can have a consistent invocation of Algorithm~\ref{alg:consensusBinary}'s $\mathsf{propose}_i(s,k)$ in $c$. Let $R$ be an execution of Algorithm~\ref{alg:consensusBinary} in which for any $c' \in R$, for any $p_i \in \sP$ we can either (i) say that $p_i$ can have consistent invocations of $\mathsf{binPropose}_i()$ in $c$, or (ii) $c'$ is the result of only consistent invocations of $\mathsf{propose}()$. In this case, we say that $R$ is a consistent execution of Algorithm~\ref{alg:consensusBinary}.

\BB \begin{theorem} 
	\label{thm:clousureConsensusBinary}
	Let $R$ be a consistent execution of Algorithm~\ref{alg:consensusBinary}. The system demonstrates in $R$ a construction of a bounded-size array of binary consensus objects.
\end{theorem} 
\BBB \begin{proof} 
	\noindent \textbf{Termination, validity, and integrity.~~} 
	Termination holds due to Theorem~\ref{thm:recoveryConsensusBinary}. Integrity holds since $p_i \in \sP$ decides by assigning a non-$\bot$ value to $O_i.est[2]$. This happens only in line~\ref{ln:Oest2GetsV} and when $O_i.est[2]=\bot$. Thus, it can happen at most once per unique pair of sequence number, $\mathit{sJ}$, and processor identifier, $\mathit{kJ}$, cf. line~\ref{ln:oAssigment} for the assignment of $O_i$'s value.
	
	With respect to validity, by line~\ref{ln:txDurbBroadcast} we can see that $\mathrm{DECIDE}(s,k,v)$ messages can only be sent with a non-$\bot$ value in the field $v$ since $\bot$ is not in the domain, $V$, of values that one can propose. Thus, when $p_i$ receives a $\mathrm{DECIDE}()$ message, line~\ref{ln:Oest2GetsV} never assigns to $O_i.est[2]$ a $\bot$-value. That is, $p_i$ decides on a non-$\bot$ value that comes from $est[1]$ of some entry $CS_j[s].BC[k]$, which in turn comes from $est[0]$ of some entry $CS_x[s].BC[k]$, where $p_j,p_x \in \sP$. Since $R$ is a consistent execution, $est[0]$ can contain only proposed values that Algorithm~\ref{alg:consensusBinary} assigns in line~\ref{ln:scdBroadcastM}. Moreover, $est[1]$ can contain only values that Algorithm~\ref{alg:consensusBinary} copied from $est[0]$ in lines~\ref{ln:est1GetsV} and~\ref{ln:nJ1OestGets}. Thus, the validity property holds.
	
	\noindent \textbf{Agreement.~~} Claim~\ref{thm:singleValue} implies agreement since it shows that only a single value can be decided in a consistent execution. 
	
	\BB \begin{claim}
		\label{thm:singleValue}
		Let $r$ be the smallest round during which any $p_i \in \sP$ URB-broadcasts $\mathrm{DECIDE}(s,k,v)$. Suppose that $p_j \in \sP$ also URB broadcasts $\mathrm{DECIDE}(s,k,v')$ during round $r$. (i) It holds that $v' = v$. Let $v''$ be the local estimate $CS_x[s \bmod M].BC[k \bmod n].est[0]$ of any $p_x \in \sP$ that proceeds to round $r+1$. (ii)  It holds that $v=v''$.
	\end{claim}
	
	\BB \begin{claimProof}{\textbf{\ref{thm:singleValue}.~~}}
		\noindent \emph{Invariant (i).~~}
		By the code of Algorithm~\ref{alg:consensusBinary}, $p_i$ receives during $r$ at least $n-t$ times the message $\mathrm{PHASE}(0, \bullet,s,k,r,v,\bullet)$, see the proof of Claim~\ref{thm:termination}. Moreover, $p_j$ has received during round $r$ at least $n-t$ times the message $\mathrm{PHASE}(1,\bullet,s,k,r,v,\bullet)$. During consistent executions, $p_x \in \sP$ can only transmit (and perhaps retransmit) one $\mathrm{PHASE}(0, \bullet,s,k,r,v,\bullet)$ message. Due to the property of majority intersection, $p_i$ and $p_j$ receive during round $r$ the same message $\mathrm{PHASE}(1,\bullet,s,k,r,w,\bullet)$ from some processor $p_x\in \sP$. Since both $p_i$ and $p_j$ executes line~\ref{ln:txDurbBroadcast} during round $r$, it must be the case that $w=v=v'$. 
		
		
		\noindent \emph{Invariant (ii).~~}
		Suppose that some correct $p_i \in \sP$ URB broadcasts $\mathrm{DECIDE}(s,k,v)$ during a round $r$. Also, $p_j \in \sP$ continues to round $r + 1$. We have to prove that $CS_j[s \bmod M].BC[k \bmod n].est[0]= v$ when $p_j$ starts round $r + 1$. Since $p_i$ URB broadcasts $\mathrm{DECIDE}(s,k,v)$ during round $r$, lines~\ref{ln:repeatPhase1End} and~\ref{ln:txDurbBroadcast} implies that there were at least $(n-t)$ nodes that have sent $\mathrm{PHASE}(1,\bullet,s,k,r,v,\bullet)$ to $p_i$ during round $r$. By the fact that $n-t > n/2$ and the majority intersection property, we know that $p_j$ also had to receive during round $r$ at least one of these $\mathrm{PHASE}(1,\bullet,s,k,r,v,\bullet)$ messages. Also, it follows from \reduce{the quasi-agreement property (}Corollary~\ref{thm:quasiAgreement}\reduce{)} that $p_j$ receives both $v$ and $\bot$ (and no other value) in the phase $1$ of round $r$, \ie $rec_j = \{v, \bot\}$, because $rec_j\neq \{v\}$ since $rec_j\neq \{v\}$ implies that $p_j$ URB broadcasts $\mathrm{DECIDE}(s,k,v)$ during $r$. Thus, $p_j$ assigns $v$ to $CS_j[s \bmod M].BC[k \bmod n].est[0]$ before continuing to round $r+1$. 
	\end{claimProof}
	
	This completes Theorem~\ref{thm:clousureConsensusBinary}'s proof.
\end{proof}

\remove{	
\Section{Self-Stabilizing Indulgent Zero-degrading Consensus} 
\label{sec:bounded}
In this section, we explain how to transform our unbounded self-stabilizing URB algorithm to a bounded one. We note the existence of several such techniques, \eg Awerbuch \etal~\cite{DBLP:conf/infocom/AwerbuchPV94}, Dolev \etal~\cite[Section 10]{DBLP:journals/corr/abs-1806-03498} and Georgiou \etal~\cite{DBLP:conf/podc/GeorgiouLS19}. The ideas presented in these papers are along the same lines. They present a transformation that takes a self-stabilizing algorithm for message-passing systems that uses unbounded operation indices and transforms it into an algorithm that uses bounded indices. The transformation uses a predefined maximum index value, say, $\mathrm{MAXINT} = 2^{64}-1$, and it has two phases. \textsf{(Phase A)} As soon as $p_i$ discovers an index that is at least $\mathrm{MAXINT}$, it disables new invocations of operations. \textsf{(Phase B)} Once all non-failing processors have finished processing their operations, the transformation uses an agreement-based global restart for initializing all system variables. After the end of the global restart, all operations are enabled. For further details, please see~\cite{DBLP:conf/infocom/AwerbuchPV94,DBLP:journals/corr/abs-1806-03498,DBLP:conf/podc/GeorgiouLS19}. 

} 

\Section{Conclusions}
\label{sec:disc}
We showed how a non-self-stabilizing algorithm for indulgent zero-degrading binary consensus by Guerraoui and Raynal~\cite{DBLP:journals/tc/GuerraouiR04} can be transformed into one that can recover after the occurrence of transient faults. We also obtained a self-stabilizing asynchronous $\Omega$ failure detector from the non-self-stabilizing construction by Most{\'{e}}faoui, Mourgaya, and Raynal~\cite{DBLP:conf/dsn/MostefaouiMR03}. As an extension, we note that Ben{-}Or~\cite{DBLP:conf/podc/Ben-Or83} presented a randomized binary consensus (using local coins). It differs from Algorithm~\ref{alg:consensusBinaryNon} only in line~\ref{ln:continueNon}, where it assigns to $est[0]$ a random binary value. This is orthogonal to the algorithm's ability to recover from transient-faults. As future work, we encourage the reader to take these building blocks into account as well as the techniques used to make them self-stabilizing when designing distributed systems that can recover from transient faults. 

\B

\end{document}